# Highest Energy Cosmic Rays and Results from the HiRes Experiment

P. Sokolsky
Physics Department
University of Utah
Salt Lake City, Utah 84112
USA

G.B. Thomson
Department of Physics and Astronomy
Rutgers - the State University of New Jersey
Piscataway, New Jersey 08854
USA

## Abstract

The status of the field of ultrahigh energy cosmic ray physics is summarized, from the point of view of the latest results of the High Resolution Fly's Eye (HiRes) Experiment. HiRes results are presented, and compared with those of the Akeno Giant Air Shower Array (AGASA), plus the Telescope Array and Pierre Auger experiments.. The HiRes measurement of the cosmic ray spectrum, and the observation of the GZK cutoff are presented. HiRes results on composition, searches for anisotropy, measurement of the proton-air total cross section, and shapes of shower profiles are presented.

**Introduction**
Experimental observations are the basis for stimulating and testing theories of cosmic ray production. At the energies of interest for this review (> $10^{17}$ eV), the main experimental observations are the energy spectrum, the chemical composition, the large and small-scale anisotropy and the correlations between them. Proposed theories about cosmic ray origins and expected propagation effects between source and detection at the Earth suggest features to look for in these observations.

The low-intensity of cosmic rays above $10^{17}$ eV requires detectors with large collecting areas. The Earth's atmosphere makes such large areas practicable because each extremely high-energy (EHE) cosmic ray produces an extensive air shower (EAS) of charged particles. The longitudinal extent of the shower is near ten kilometers of atmosphere while the footprint of charged particles on the ground has a radius of order of a kilometer. Thus the showers' characteristics can be determined by sampling the particle densities with a giant array of widely spaced particle detectors on the ground, or by detecting air-fluorescence light produced by the passage of the shower particles through the atmosphere. Contemporary giant ground arrays have collecting areas in the range of 100 $km^2$ (AGASA) [1] to thousands of $km^2$ (Telescope Array [2] and Pierre Auger [3] experiments). The Fly's Eye air-fluorescence detector [4] had an effective time averaged aperture of approximately 100 $km^2$str at energies near $10^{20}$ eV while the High Resolution Fly's Eye (HiRes) [5], Telescope Array and Auger air-fluorescence detectors have order of magnitude higher time-averaged collecting areas. These detectors can only operate on clear, moonless nights and thus have a duty factor of near 10%.

An EAS measurement provides indirect information about the primary particle. To advance understanding of the origin of EHE cosmic rays, what is needed is the energy, atomic mass and arrival direction for each particle. The indirect measurements made necessary by the low flux of these particles makes acquiring such information difficult. Shower measurements are always incomplete. Ground arrays sample particle densities sparsely and at one altitude only. The stochastic nature of the shower guarantees that particle densities will fluctuate around their expected values for a cosmic ray of a certain energy and mass. Thus even a perfect measurement would not uniquely determine the properties of the cosmic ray particle that initiated the EAS. Furthermore, ground array measurements must rely on a model of high-energy interactions to compute the expected particle densities for a cosmic ray of a particular mass and energy. Any such model is an extrapolation from known interaction properties at lower center-of-mass energies. Uncertainty in the hadronic interaction model implies an uncertainty in the expected air shower footprint for a particle of a particular mass and energy.

The air-fluorescence measurement is much less dependent on extrapolation and gives a more reliable estimate of the primary shower energy, as it is inherently calorimetric. The magnitude of the air-fluorescence is simply proportional to the energy deposited in the atmosphere by the EAS. However, each detection technique encounters different manifestation of these intrinsic sampling difficulties and other problems associated with the experimental technique itself. The most recent experiments (Telescope Array and Auger) have deployed both kinds of detectors in the belief that "hybrid" observation will reduce the difficulties previously encountered.

While the pioneering ground arrays of Volcano Ranch [6], SUGAR [7], Haverah Park [8], Yakutsk [9] and Akeno [10] made many important contributions to our understanding of the physics of cosmic rays above $10^{17}$ eV, this review will concentrate on the more recent results: AGASA (a ground array of plastic scintillator detectors) and HiRes (the second generation Fly's Eye air fluorescence detectors). In addition preliminary results from the Pierre Auger experiment will be discussed and the status of the Telescope Array experiment, currently under construction, will be examined.

**Theoretical Background**

Theoretical expectations of what one should see in the ultrahigh energy regime (above $10^{17}$ eV) include the GZK cutoff. First predicted by K. Greissen, G. Zatsepin, and V. Kuzmin in 1966 [11] this is a high-energy limit to the flux of cosmic rays. It is caused by interactions between cosmic ray protons of extragalactic origin and photons of the Cosmic Microwave Background Radiation (CMBR), where the center of mass energy is above the threshold for pion production. Just above this threshold lies the $\Delta(1232)$ resonance that has a quite large cross section. The pions from the decay of this resonance carry away about 20% of the proton's energy. This is thus a strong energy-loss mechanism for protons, and repeated photopion interactions decrease a proton's energy until it falls below inelastic threshold. The threshold is set by the temperature of the CMBR and the mass and width of the $\Delta(1232)$ resonance. It is about $6 \times 10^{19}$ eV if extragalactic sources are uniformly distributed throughout the universe. If there were a local overabundance or under-abundance of sources the "GZK energy" could change by a small amount, but the effect would be evident mostly above $10^{20}$ eV [12]. Thus the shape of the spectrum fall-off above the GZK cutoff should tell us about the local abundance of sources. A steep fall-off would indicate an under abundance of sources in the local area. Finally, the distance protons must travel before their energy is brought below the inelastic threshold is set by the $\Delta(1232)$ cross section and the density of CMBR photons, and is about 50 Mpc.

Electron-positron pair production also occurs in the interactions between protons and CMBR photons. Much less energy is lost by the proton in such an interaction, so it takes more than an order of magnitude longer travel distance before there is a noticeable effect on the proton's energy. This energy loss mechanism is predicted to excavate a dip in the spectrum in about the middle of the $10^{18}$ eV decade [13]. As discussed below, data shows that there is indeed a dip in the spectrum at this location, which is called the "ankle". But whether the ankle is caused by pair production is still disputed. The main other predicted cause of the ankle is the galactic/extragalactic transition. The fall to the ankle is here seen as the last gasp of the galactic cosmic ray flux, and the rise above the ankle is the extragalactic flux showing through.

One can use other data to determine which of these two explanations of the cause of the ankle is correct. If the ankle is caused by electron-positron pair production then the cosmic rays in this energy range are predominantly extragalactic and their composition should be uniformly light since photospallation will break up heavy nuclei over these distances. For a variety of reasons including leakage from the magnetic containment of the galaxy and preferential acceleration of heavy nuclei at the acceleration source, galactic cosmic rays are expected to be dominated by heavy nuclei near the end of the galactic spectrum. Thus, in the case of the galactic – extragalactic transition hypothesis, a heavy galactic composition should give way to a light extragalactic one and a change of composition should be evident. We shall see that the composition measurements made by the HiRes prototype (operating in hybrid mode with the MIA muon array) and by HiRes stereo indicate an answer to this question. These data show that the composition is approximately constant over the whole ankle region, indicating that the ankle is not the galactic-extragalactic transition.

In performing fits to the cosmic ray spectrum ankle [14] one sees that the spectral indices below and above the ankle are sensitive to two important parameters about extragalactic cosmic ray sources, the power law index at the source and the evolution of the sources. Thus it is important to measure the properties of the ankle accurately.

It is commonly accepted that galactic sources of cosmic rays are supernova remnants. The Kascade experiment [15] has shown that there is a rigidity-dependant cutoff at and above the "knee" of the spectrum, which occurs for protons at about $3 \times 10^{15}$ eV which corresponds approximately to estimates of the maximum energy of acceleration for these remnants. Above this energy the composition gets heavier until about $1 \times 10^{17}$ eV, when the limit is reached for iron primaries. To explain the origin of cosmic rays of even higher energies one must invoke other source types, such as supermassive black holes in distant galaxies, known as AGN's [12] (the supermassive black hole at the center of the Milky Way galaxy is relatively small and inactive). These can accelerate cosmic rays to higher energies. These are not likely to be sources where a great deal of fusion of heavy elements has occurred, and any nuclei that are accelerated are broken up by spallation, so extragalactic cosmic rays are expected to be of light composition (hydrogen and helium nuclei).

In addition to these three predicted sources of structure in the cosmic ray spectrum and composition, there is an observed feature that is not predicted, called the "second knee." This is located in the middle of the $10^{17}$ eV decade and is a steepening of the spectral falloff (from about $E^{-3.0}$ to $E^{-3.3}$ spectral behavior). Because of differing energy scales among the various experiments that have seen it, the exact energy of the second knee is not known. Collecting data on spectrum and composition in the energy region of the second knee should be of the highest priority for future experiments.

This theoretical and experimental situation raises several questions to be answered by the current generation of experiments.

1. Is the GZK cutoff present? At what energy does it occur? How steep is the falloff above the cutoff? Is there a local under or over abundance of sources?
2. What is the energy of the ankle? What are the power law indices below and above it?
3. What is the composition as a function of energy? Is it heavy or light? Can we see a transition region? Is the transition at the ankle or lower in energy?
4. What is the cause of the second knee? Is the second knee a galactic feature (if so, it would show up in the heavy component of the flux) or an extragalactic feature (the composition would then be light)?
5. What are the sources of the highest energy cosmic rays? Is there anisotropy in the cosmic ray flux that points to sources? Is there large-scale anisotropy related to galactic magnetic fields?

**History**
The AGASA experiment was an expansion of the Akeno ground array, located at the Akeno observatory in Japan (35.8deg N 138.5 deg E). The AGASA array consisted of 111 plastic scintillators each with 2.2 $m^2$ area spaced by approximately 1 km and covering an area of ~ 100 $km^2$. These scintillators have an essentially equal response to the electrons and muons in the EAS. In addition there were 27 sets of proportional counters located under absorbers, which were used to measure the muon component of the EAS. The muon threshold energy was .5 GeV. Full operation of the array began in 1993 and it was decommissioned in 2004. All the counters were connected by an optical fiber network.

The Fly's Eye experiment was located at Dugway, Utah (40deg N, 113 deg W) at an atmospheric depth of 860 $gm/cm^2$. There were two detectors, Fly's Eye I consisting of 67 spherical mirrors of 1.5 m diameter and Fly's Eye II, consisting of 36 similar mirrors located 3.3 km away. Each phototube subtended a 5 by 5 degree pixel in the sky. Events were observed in monocular mode when either detector satisfied a minimum trigger requirement and in stereo mode when both detectors triggered on the same event. The monocular time-averaged aperture approached 100 $km^2$str near $10^{20}$ eV while the stereo aperture was ~40 $km^2$ str above $10^{19}$ eV. The detectors were operated from 1982 to 1992.

A second-generation air fluorescence experiment, the High Resolution Fly's Eye (HiRes) was proposed in the early 1990's and was completed in 1997 (HiRes I) and 1999 (HiRes II). It was also located at Dugway, Utah and had two air-fluorescence detector sites, HiRes I and HiRes II spaced 12.6 km apart. Dugway Proving Grounds has a clean atmosphere and low light pollution. Each HiRes site was on a hill, above the bulk of most of the aerosol haze in the atmosphere. The two sites gathered data independently and the data was analyzed in either monocular or stereo mode. This detector had smaller phototubes resulting in a pixel size of 1 degree by 1 degree on the sky. Among other improvement over the original Fly's Eye was an FADC data acquisition system at HiRes II which allowed a much more precise measurement of the longitudinal shower profile. The HiRes I detector took data in monocular mode from 1997 to 2006, while HiRes II operated from 1999 to 2006. Consequently the data set from this experiment is divided into a large monocular data set from HiRes I, a monocular data set from HiRes II and a combined stereo data set covering 1999 to 2006.

**The Air Fluorescence Technique**
Ionizing particles traversing the atmosphere produce light by exciting the 2P and 1N band systems of N2 and $N2^+$ molecules [16]. This excitation and the resultant fluorescence have been measured by a number of groups [17] as a function of pressure and temperature, most recently by the FLASH [18] and Air Fly [19] collaborations. In the 300 to 400 nm wavelength region that is effective in producing photomultiplier signals the air-fluorescence efficiency is approximately 4.0 photons/m per ionizing particle. The competition between fluorescence and collisional de-excitation makes this fluorescent yield nearly independent of atmospheric pressure (up to ~ 12 km altitude).

As an EAS propagates through the atmosphere, its charged particles excite air-fluorescence. This light is emitted isotropically. In addition to air fluorescence, Cherenkov light is produced within a cone of approximately 2.5 degrees of the charged particles. The resultant Cherenkov light distribution, reflecting the multiple scattering of electrons, is strongly beamed forward along the EAS axis. Air fluorescence detector mirrors gather these photons and focus them onto arrays of photomultiplier tubes (PMTs). Each PMT views a small solid angle of the sky. An EAS produces a characteristic pattern of pixel triggers as it propagates through the atmosphere with nearly the speed of light. Air fluorescence detectors record the PMT signal amplitude as a function of time (set by either the phototube crossing time or the FADC clock) and this information, after appropriate corrections, can be used to determine the longitudinal shower profile (number of ionizing particles as a function of atmospheric depth). The integral of this profile is proportional to the primary shower energy and the proportionality constant is largely independent of the hadronic physics at the primary interaction.

Photons from an EAS travel through the atmosphere to reach the mirrors and will scatter on the air molecules and aerosols in their path. Molecular or Rayleigh scattering is well understood and the effect is straightforward to calculate once the molecular density profile is known (for example from balloon radiosonde measurements). Scattering from aerosols varies with the aerosol content of the air and the nature of the aerosols and must be measured on an hourly basis. The effect of aerosol scattering is typically characterized by the horizontal aerosol extinction length, the exponential drop-off of aerosol density with altitude, the vertical aerosol optical depth, and the angular dependence of the scattering (the phase function). Lasers with steerable beams, located at some distance from the detector sites are used to sweep through the air-fluorescence detector aperture and determine these parameters [20]. In clear desert atmospheres such as are present at the Dugway site, the net effect of aerosol variations on the total attenuation of light is small and is well approximated, over the course of an experiment, by an average stable atmosphere.

Lasers also give an indication of the presence of clouds in the atmosphere since a laser beam hitting a cloud will mushroom out due to enhanced multiple scattering. Other tools for measuring the atmosphere include infrared cloud monitors and vertically directed xenon flashers located between the air-fluorescence detectors [21].

Air fluorescence detectors need to be carefully calibrated since the energy of an EAS, while calorimetrically determined, depends on the absolute gain of the detector. Calibration is typically done using NIST- traceable stable light sources and an overall end-to-end calibration is available using laser beams whose energy at injection into the atmosphere is carefully measured using NIST calibrated detectors [22]. The relation between PMT signals and this energy give the overall detector calibration constant. Typical systematic errors for this absolute detector calibration are 10%.

**Event Reconstruction**

The reconstruction of the EAS longitudinal shower profile depends on the determination of the shower geometry, i.e. the impact parameter of the shower ($R_p$) and the shower zenith and azimuth angle. This can be determined in monocular mode by fitting the expected dependence of the relative arrival time of the signals in the sequential pixels in the shower-detector plane on these parameters. The shower-detector plane is determined by the best fit to a plane subtending the direction vectors of the pixels triggered by the EAS. When stereo information is available, the shower geometry is simply the intersection of the two shower detector planes (from HiRes I and II, for instance). Once the geometry is known, corrections for solid angle effects, variations in acceptance from pixel to pixel (due to shadowing of the mirrors by the phototube cluster), mirror reflectivity and PMT and electronics gain as well as atmospheric attenuation are applied. The net result is a measurement of the number of photons emitted at the EAS location as a function of atmospheric depth in bins dictated by either the pixel size (one degree for HiRes) or by the FADC timing slice (typically 100 ns). Knowledge of the air-fluorescence efficiency then allows conversion to shower size (number of ionizing particles) versus atmospheric depth.

**The Ground Array Technique**

Ground arrays sample the number of particles at observation level in an EAS produced by a primary cosmic ray. There are significant fluctuations in the position of depth of shower maximum ($X_{max}$) for an event with the same energy and atomic mass. Hence, simply summing the total density at observation level is inadequate. Hillas [23] proposed a method to much reduce the influence of shower development fluctuations on the determination of shower energy by ground array experiments. The method relies on the fact that particles at distances of 500 to 1000m from the core of the EAS at observation level are most likely to have their origin at shower maximum. Since the number of particles at shower maximum is proportional to the initial particle energy and the fluctuations in particle number are minimized at this stage of shower development, a measurement of shower particle density far from the core should give more accurate results. Hillas' idea has been confirmed by many simulations and has been incorporated in the analysis of data by the Haverah Park [24] and Yakutsk [25] groups in the past. While the details depend on the hadronic model used and on the nature of the primary particle, all simulations agree that the fluctuations are minimized near 600 m for EAS with energies near $10^{19}$ eV near sea level. A complication comes from the fact that showers at non-zero zenith angles have different degrees of attenuation in the atmosphere. Hence the distant density measurement $\rho(600)$ has to be corrected for a zenith angle related effect. This correction factor can be determined from the simulations but is then hadronic model dependent. It can in principle also be determined experimentally by using the equal intensity method. This posits that since the cosmic ray flux impacting the atmosphere at a given energy should be independent of zenith angle, the dependence of cosmic ray rate on zenith angle for a fixed interval in $\rho(600)$ can be used to derive this correction factor. An additional assumption that must be made is that the composition of cosmic rays is not changing rapidly in the energy region of interest. The lateral distribution of charged particles, especially the muon component, will be different near 600m for p and Fe incident nuclei, for example. The extensive air showers produced by a heavy composition will also attenuate much faster with zenith angle than those from a light composition. Additionally, this method requires significant statistics and, given the power law nature of the cosmic ray flux, is difficult to use at the highest energies.

Hybrid experiments such as the Telescope Array and Auger have another option. The energy scale obtained by using the $\rho(600)$ parameter and its zenith angle dependence can be calibrated using events where the air-fluorescence detector also has a good energy measurement. This largely removes any dependence on hadronic models and complex simulation codes. However, since the on time for air-fluorescence detectors is usually less than 10% of the ground-array on time, the calibration becomes uncertain at the highest energies due to limited statistics. Currently, linear extrapolation is used to avoid this problem [26], but this moots the question of whether the zenith angle dependent correction factors for the ground array are in fact linear with energy. Hence the systematics of ground array calibrations, either using simulations or calibrations from air-fluorescence detectors will remain problematic at the highest energies of interest.

**Hadronic Interaction Models and EAS Simulation**

The measurable aspects of an EAS (Energy, Xmax, lateral and longitudinal profile, number of electrons or electron density and number of muons or muon density) depend on the nature of the primary particle and the hadronic interaction physics at each interaction and decay point. Since the CM energy of the primary interaction is well beyond accelerator energies, theoretical models must be used to predict multiplicity, inelasticity, the total inelastic cross-section and other variables. Nuclear effects are also important, particularly if the incident particle is a heavy nucleus. Several QCD based theoretical models such as DPMJET [27], nexus [28], QGSJET [29] and SIBYLL [30], are widely used to predict EAS development. These models differ in their predictions of a number of parameters. The total inelastic cross-section in particular differs significantly in these models at the highest energy (see figure 14). An independent measurement of this quantity at ultra-high energies can serve as a test of these models and as a step to achieving a self-consistent description of the EAS development as well as being of great interest to high-energy physics. Methods for determining this are discussed below.

While the distribution of $X_{max}$ in the atmosphere depends almost entirely on the nature of the primary particle and the choice of interaction model for the first few generations in an EAS, the situation is different for the lateral distribution for electrons and muons, particularly at large distances from the shower core. Since the average energy of the parent pion or kaon that produces muons at a kilometer from the core is on the order of 100 GeV, it is the low-energy part of the EAS simulation that is also an issue [31]. Since the parent hadrons are typically produced at small Feynman x where there is almost no accelerator data, theoretical extrapolations must be used even in this energy range. Figure 1 shows the surprising result: the various low-energy simulation packages give significantly different predictions for the muon density $\rho_\mu(1000)$ and even the electron density [32].

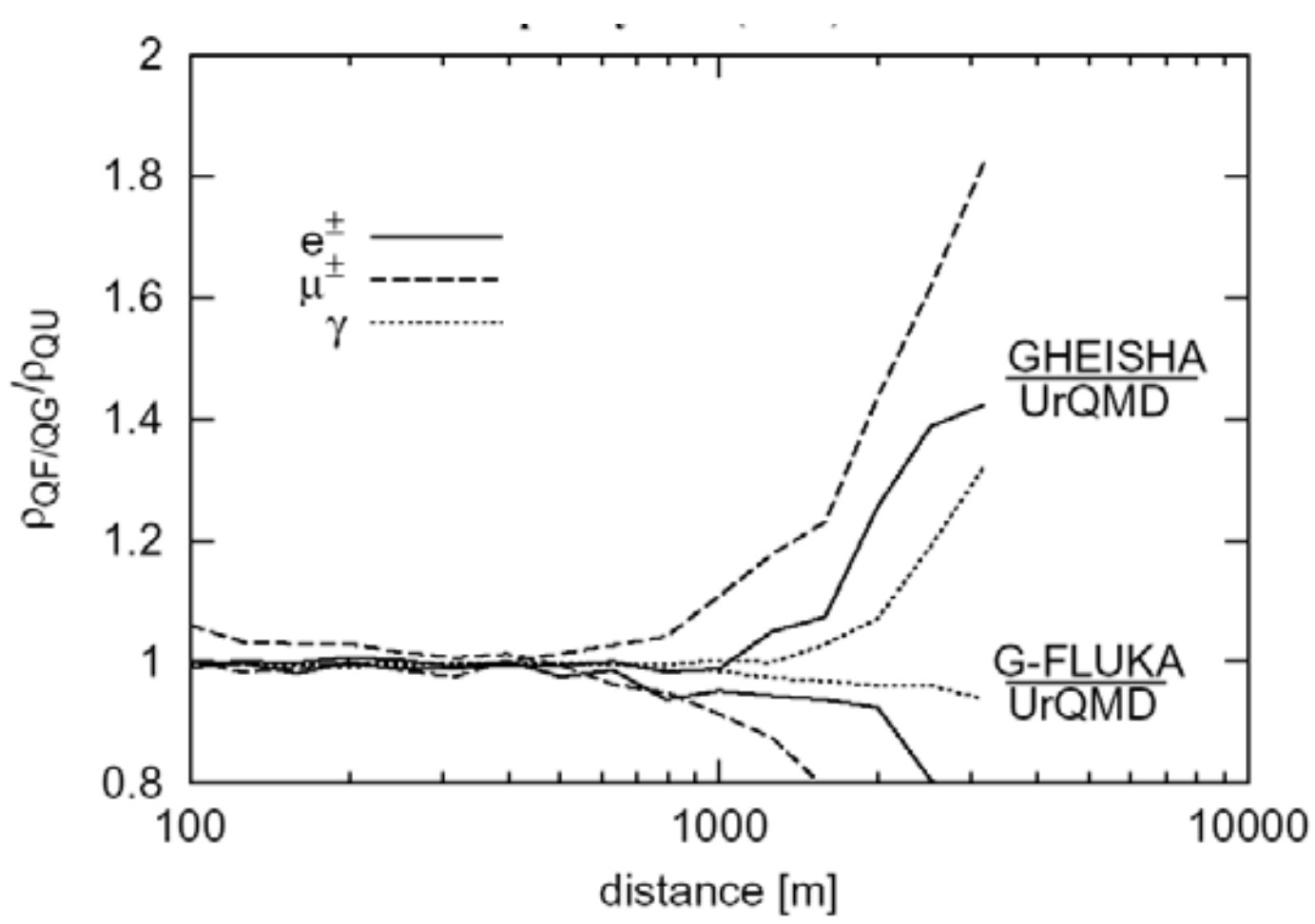

Figure 1. Dependence of lateral distribution functions on lower-energy models. The high-energy part of the shower is simulated using QGSJet while GHEISHA, G-FLUKA and UrQMD are used as low energy models.

A number of experiments have been proposed or are in the planning and implementation stage, which will hopefully alleviate this issue as well as improving our knowledge of the higher energy physics. The Fermilab MIPP experiment [33] will cover the entire range in Feynman x for a variety of beam and secondary particles near 100 GeV. Proposed experiments at the LHC [34] will enrich our knowledge of small-x physics at the highest energies and will further constrain hadronic models used in EAS simulations.

A related issue is the treatment of baryons in the EAS. Data from RHIC on parton saturation effects [35] have only recently been incorporated into EAS simulation models [36] (see below). A correct accounting of these effects may in fact be the key to a self-consistent description of EAS in terms of lateral and longitudinal development.

**Fly's Eye and AGASA**

The Fly's Eye and AGASA spectrum measurements set the stage for those made by HiRes. The Fly's Eye experiment made two types of spectrum measurements, monocular and stereo [37]. The two differed in the ability to observe the "ankle" of the cosmic ray spectrum. It is thought that the better resolution of the Fly's Eye stereo spectrum made the difference in this observation. The aperture of this pioneering experiment was a bit too small to measure the spectrum at $10^{20}$ eV, hence too small to observe the GZK cutoff. However one event was collected, at an energy of $3.2 \times 10^{20}$ eV, well above the expected cutoff energy. This was a remarkable event, only 12 km from the Fly's Eye I detector, which made a good measurement of its profile and energy [38]. Unfortunately, the event fell behind the Fly's Eye II detector so was not seen in stereo. This is one of a set of events, seen by experiments too small to measure the spectrum at the GZK energy that seemed to predict that something unusual was occurring in the GZK energy region [39].

The AGASA experiment was the first to have a large enough aperture to measure the spectrum at the GZK energy with reasonable statistics [40]. Figure 3 shows their result (inverted triangles) where the y-axis is the differential flux multiplied by $E^3$ This transformation removes the overall power-law trend and allows details in the spectrum to be seen more clearly. Their result was different from the Fly's Eye stereo spectrum in that the spectrum was about a factor of 2 higher in normalization. This could be due to a different energy scale in the two experiments. Another aspect of their spectrum was very startling: they saw the spectrum continuing as if the cutoff were absent. The most significant part of their result was the observation of 11 events above $10^{20}$ eV. Their measurement of the spectrum launched many theoretical speculations as to how the GZK cutoff mechanism might possibly be evaded.

However their result was retracted in 2006 [41]. They discovered an error in their data analysis having to do with the zenith angle correction (see above). The effect of improving their analysis was to lower their energy scale by about 10%, and correct an energy scale nonlinearity which lowered their highest energy events by about 15%. The result was reducing the 11 events above $10^{20}$ eV to 5 or 6. The AGASA collaboration also dropped their claim that the GZK cutoff is absent, saying instead that their data do not have the statistical power to make a claim either way.

**The HiRes Monocular Spectrum**

The HiRes experiment, on the other hand, has observed the GZK cutoff. Using the fluorescence experimental and analysis techniques described above plus an augmented Monte Carlo simulation method, the spectrum was measured in monocular mode and it shows the GZK cutoff at the 5 $\sigma$ level.

An important aspect of measuring the spectrum with a fluorescence experiment is calculating the aperture. Since that aperture varies with energy (higher energy events are brighter and hence can be seen at larger distance) one might expect that the aperture calculation by the Monte Carlo (MC) technique would be more difficult than for ground array experiments, but just the reverse is the case.

The Monte Carlo technique in cosmic ray physics has two parts: simulation of showers using a shower MC program such as Corsika [42] or Aires [43] and hadronic generator routines (to extrapolate cross sections measured by accelerator experiments to higher energies) such as QGSjet [29] or Sibyll [30]. The second part is a detector simulation program where the air-fluorescence light generated by the showers is propagated to the apparatus and the detector response calculated, or, for a ground array, the number of particles that strike a detector is translated into a pulse height.

The shower MC process has several difficult points where ground arrays are concerned. First, following every particle in the shower is prohibitive in time (a $10^{20}$ eV shower takes a month to generate on a typical computer), so methods have been developed to speed up the generation of showers by following particles statistically, but they lose information on the fluctuations in shower development in the tails of showers. Since ground arrays work only in the tails of the lateral distribution (the central part of a shower would also drive a ground array detector into saturation) one cannot reliably predict the resolution of a ground array experiment by this method. In addition the uncertainty in the shower simulation process due to extrapolation of cross sections in energy makes the entire simulation process model dependent. Even calculating the efficiency of a ground array reliably by this technique may be beyond the state of the art.

On the other hand, for a fluorescence experiment that sees only the central part of the shower, where there may be $10^{11}$ particles at shower maximum, calculating shower development statistically is straightforward. In addition, the energy measurement is done calorimetrically, so a fluorescence experiment is insensitive to cross section extrapolations. Thus it should be possible to perform an accurate aperture calculation for a fluorescence experiment.

The largest effect on the aperture (in addition to the intrinsic brightness of showers) comes from the fact that fluorescence detectors designed for high-energy observations cover a limited extent of the sky in elevation. HiRes and TA detectors look from 3 to 31 degrees in elevation, where Auger detectors look up to 29 degrees. This bias affects the aperture and to perform an accurate MC detector simulation one must make sure the events in the simulation populate the correct angular range in elevation. One does this by making sure they have the same $X_{max}$ distribution as the data. Generating a correct mix of proton and iron events from a given hadronic model generator, for example QGSjet, can mimic the measured $X_{max}$ distribution well (it turns out to be about an 80/20 ratio for protons/iron above $10^{18}$ eV for QGSjet). It does not matter if QGSjet correctly simulates protons or iron for the aperture calculation as long as it represents the measured distributions adequately. The same is true of Sibyll (here one must use about a 60/40 ratio). In fact, one can use either hadronic generator program and one gets the same answer for the experiment's aperture to within a few percent. Thus for a fluorescence experiment one can simulate the aperture and measure the spectrum in a model-independent way.

To calculate the aperture as a function of cosmic ray energy, a very accurate Monte Carlo simulation of the HiRes experiment was performed [44]. Two libraries of cosmic ray showers were generated using the Corsika shower program and QGSJet hadronic generator, one for proton and one for iron primaries. A detector simulation program placed library events in the atmosphere in the vicinity of the HiRes-II detector, calculated the fluorescence and Cerenkov light generated by the showers, and how much light would have been detected. A complete simulation of the optical path, trigger, and readout electronics was performed. This simulation followed the experimental conditions that pertained over the data-collection period. The result was written out in the same format as the data and analyzed by the same programs. The stereoscopic energy measurement of the Fly's Eye experiment [37], and the HiRes stereo Xmax composition measurement [48] were used as inputs.

The Monte Carlo simulation can be checked for accuracy in representing the detector trigger efficiency by comparing many Monte Carlo distributions of geometrical and kinematic variables to the data. The comparisons are in excellent agreement and indicate good understanding of how to simulate the experiment.

The result of the monocular spectrum measurement is shown in figure 2. In this figure the ankle is seen clearly as a dip in the spectrum centered at log(E)=18.6, and a sharp fall- off is seen at about log(E)=19.8, which is $6\times10^{19}$ eV, exactly the expected GZK energy. In figure 2 a set of three lines is shown, representing a multi power law fit to the data. A fit with a single power law over the entire energy range has a $\chi^2$ of 162 for 39 degrees of freedom. A fit using two power laws with one break allowed (and fit by MINUIT) finds the break at the ankle at log(E)=18.65±0.05, and has a $\chi^2$ of 62.9 for 37 degrees of freedom (dof). If two break points are allowed, as shown in the figure, they are found to be at 18.65±0.05 and at 19.75±0.04. In other words the data show the GZK energy to be $10^{19.75}$ eV. The fit has a $\chi^2$ of 39.5 for 35 dof. The improvement in $\chi^2$ for a change of 2 dof implies that the two break point fit is preferred at a confidence level corresponding to $4.5\sigma$.

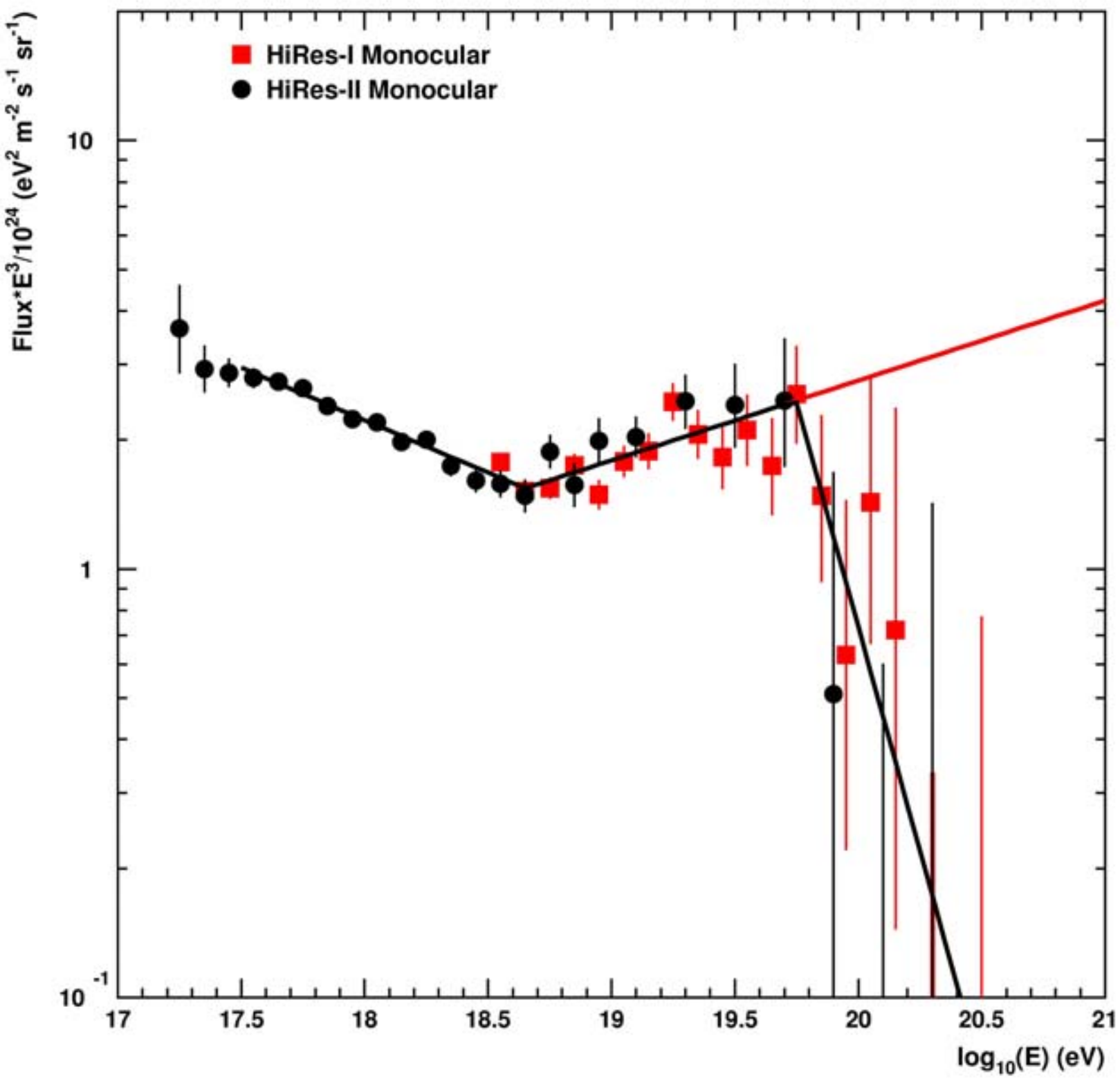

Figure 2. Spectrum measured by the HiRes detectors in monocular mode. Red points are the spectrum from HiRes-I and black are from HiRes-II. The ankle and the GZK cutoff are evident in these data. The black line is a fit to the data described in the text, and the red line is an extension of the black line segment (from the ankle to the GZK cutoff) extended to higher energies.

One can also quantify the significance of the break by extending the fit rising from the ankle and comparing the number of events one would expect to see, if the break were absent, with the number we actually see. The numbers are 39.9 expected and 13 seen. The Poisson probability for the observed deficit corresponds to a significance of 4.8 $\sigma$, consistent with the previous estimate. This estimate takes into account that some events are seen in both HiRes-I and HiRes-II mono datasets, and removes the overlapping events from the HiRes-I mono dataset.

The break in the spectrum is definitely present. The break occurs at the energy predicted for the GZK cutoff. One therefore concludes that it is the GZK cutoff. The spectral indices to the left and right of the break are 2.81±0.03 and 5.1±0.7. The fall-off to the right of the break is very steep, which may have implications for the over or under abundance of sources in the local area (within about 50 Mpc). Figure 3 shows the HiRes mono spectra compared with previous fluorescence experiments' results: the Fly's Eye stereo spectrum and the HiRes/MIA hybrid spectrum. These three measurements are in good agreement. The figure also shows the same HiRes mono spectra plus the original AGASA spectrum.

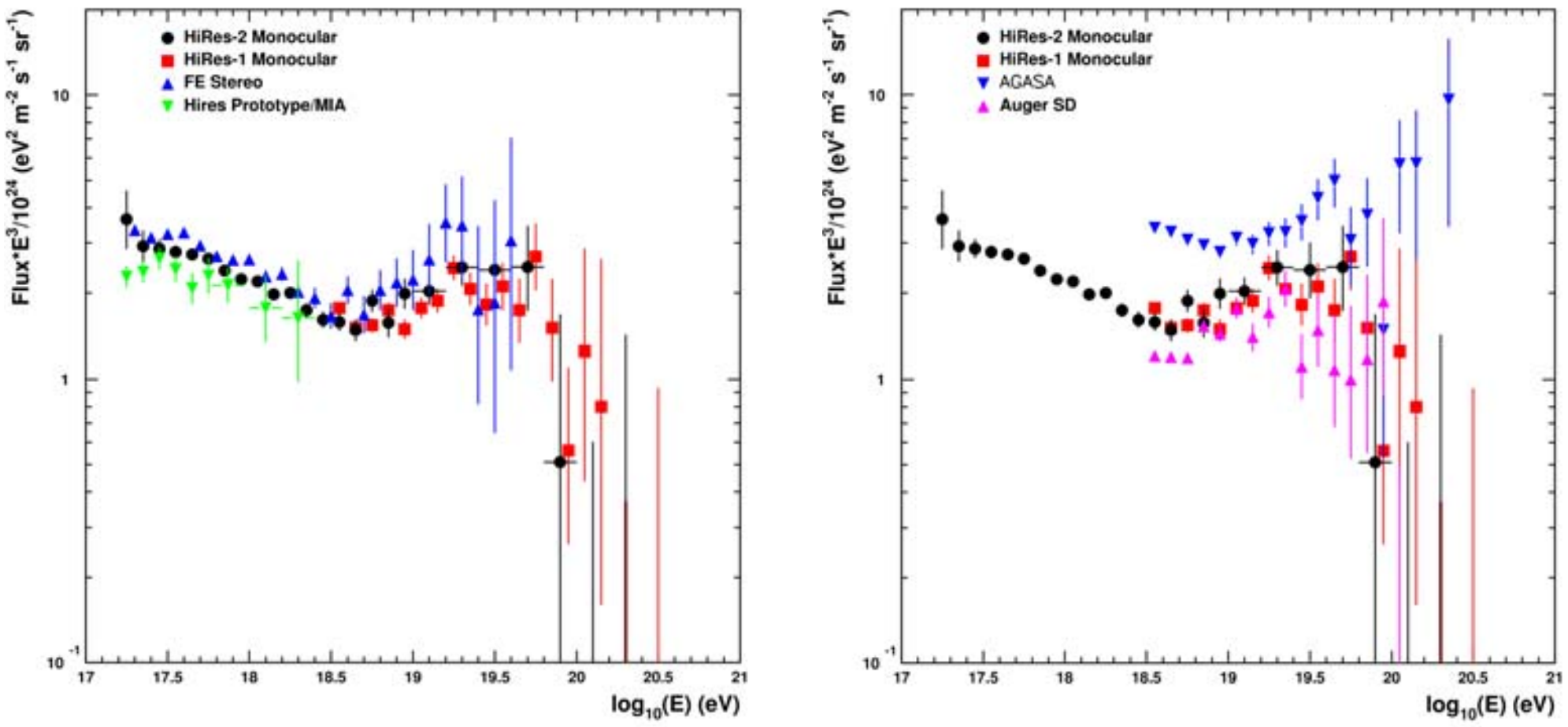

Figure 3. The left part shows the two HiRes monocular spectra in red and black. In addition the Fly's Eye stereo spectrum is shown in blue and the HiRes/MIA spectrum is green. The right part shows the two HiRes monocular spectra in red and black, plus the original AGASA spectrum in blue and the 2005 Auger spectrum in magenta.

The HiRes stereo spectrum is shown in figure 7. This spectrum is consistent with the monocular spectra and also shows the ankle and break due to the GZK cutoff. With two detectors observing the same events one can use the two reconstructions to measure the energy resolution. This is shown in figure 5, where (E1-E2)/E2 is shown (here 1 and 2 refer to the two HiRes detectors). The result is approximately gaussian and is consistent with resolutions of about 30% and 15% for HiRes1 and HiRes2 respectively at $10^{18.2}$ eV.

**The HiRes Stereo Data**

The HiRes experiment was designed as a stereo experiment and while the monocular data has the largest statistics because of the early turn-on of HiRes I, the stereo data set has the best geometric, $X_{max}$, and energy resolution. The expected $X_{max}$ and energy resolutions are shown in figures 4 and 5. This is based on simulated events as described in the composition section. As indicated there, stereo data allows two independent measurements of the same event and the "pull" distribution for data and simulations should be similar if the detector resolution is correctly estimated. Figure 8 show that this is indeed the case.

The calculation of an energy spectrum requires a good understanding of the detector aperture. Since the aperture for an air-fluorescence detector grows with energy and the aperture of a stereo detector is the intersection of two growing apertures, careful crosschecks are necessary to insure that the calculation is correct.

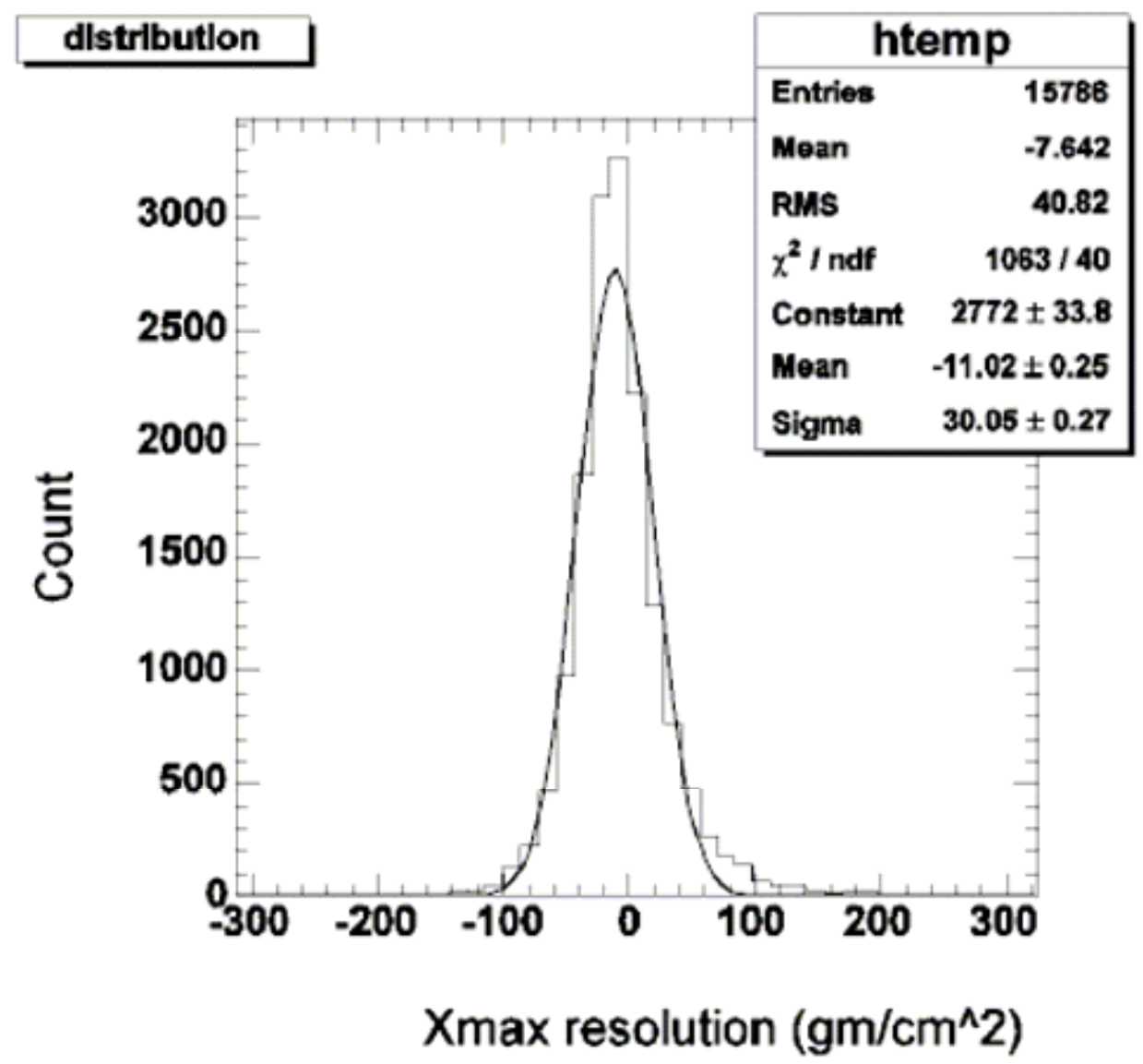

Figure 4. Stereo Hires Xmax resolution.

The geometric distribution of showers reconstructed by the HiRes experiment reflect precisely the detection efficiency of showers and can be used to check the efficacy of the simulation. All distributions in geometrical variables for data and for simulations, such as the impact parameter Rp, zenith and azimuth angles are in good agreement.

The HiRes stereo aperture is shown in Fig. 6. The aperture varies very rapidly below 3 x $10^{18}$ eV, then flattens out and saturates at near 10,000 $km^2$str at the highest energies. Since about half of the stereo data comes from distances where the detector trigger efficiency is falling off, it is useful to also define a "fully efficient" aperture. Here one sets a maximum impact parameter for each energy bin, where the value of $R_{pmax}$ is determined as the distance at which the trigger efficiency just begins to decrease. The "fully efficient" aperture is also shown in figure 6. Note that most of the difference between the two apertures is at low energies. Defining a "fully efficient" aperture has the added benefit that the resultant spectrum is much less sensitive to atmospheric extinction variation. Changes in atmospheric extinction do not affect the triggering efficiency of events that are close enough to the detector.

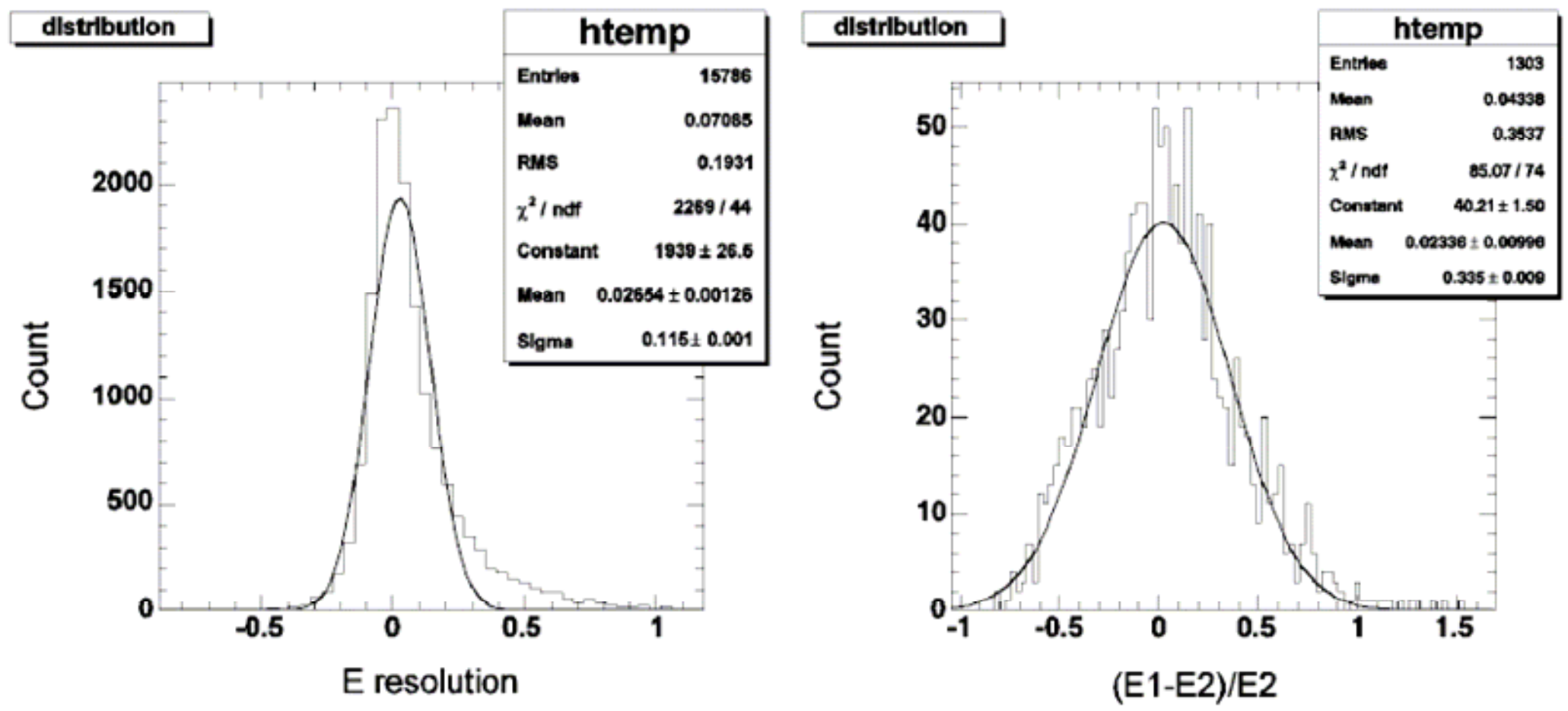

Figure 5. HiRes stereo energy resolution (left) and energy difference distribution (right).

After cuts to make sure that the shower profile is well-measured figure 7 shows the spectrum for the "fully efficient" aperture [45]. No cloud cuts have yet been applied to this spectrum, however. Structure in the spectra and spectral normalization are in good agreement with monocular data.

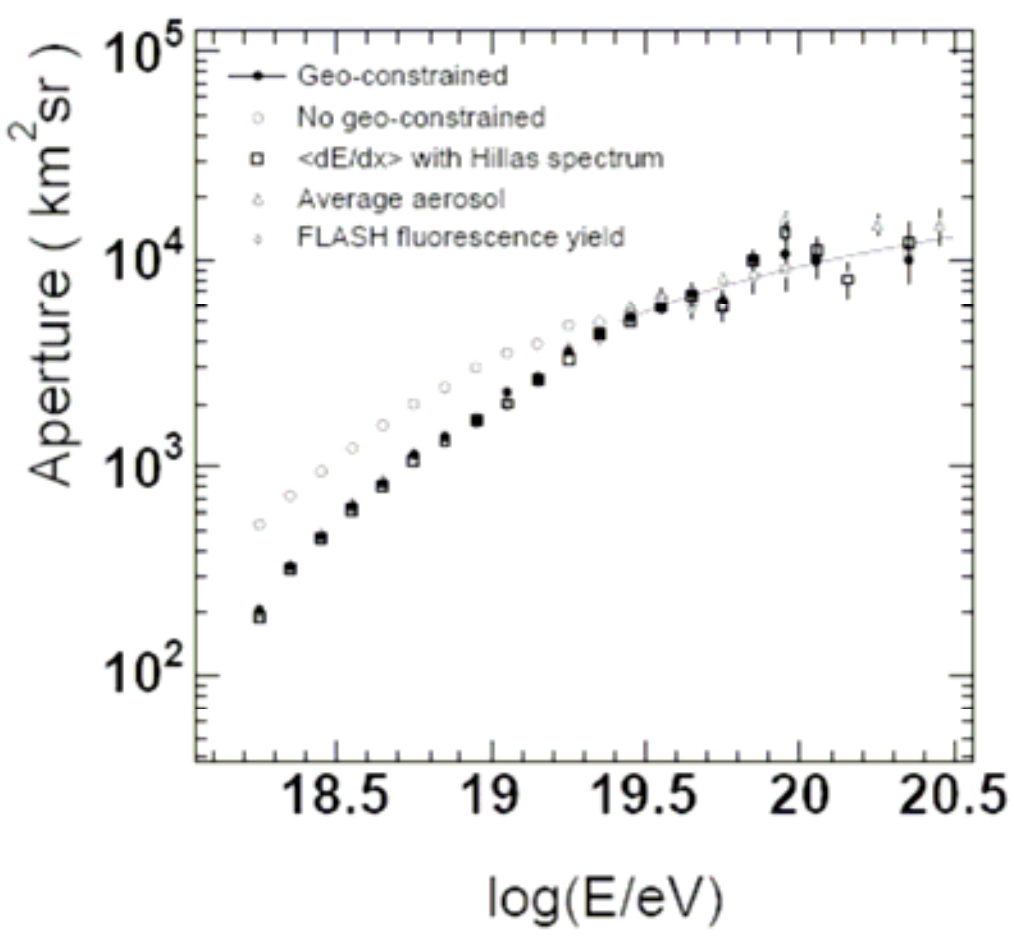

Figure 6. HiRes stereo aperture. Open symbols are the total aperture. Filled symbols represent the "fully-efficient" aperture.

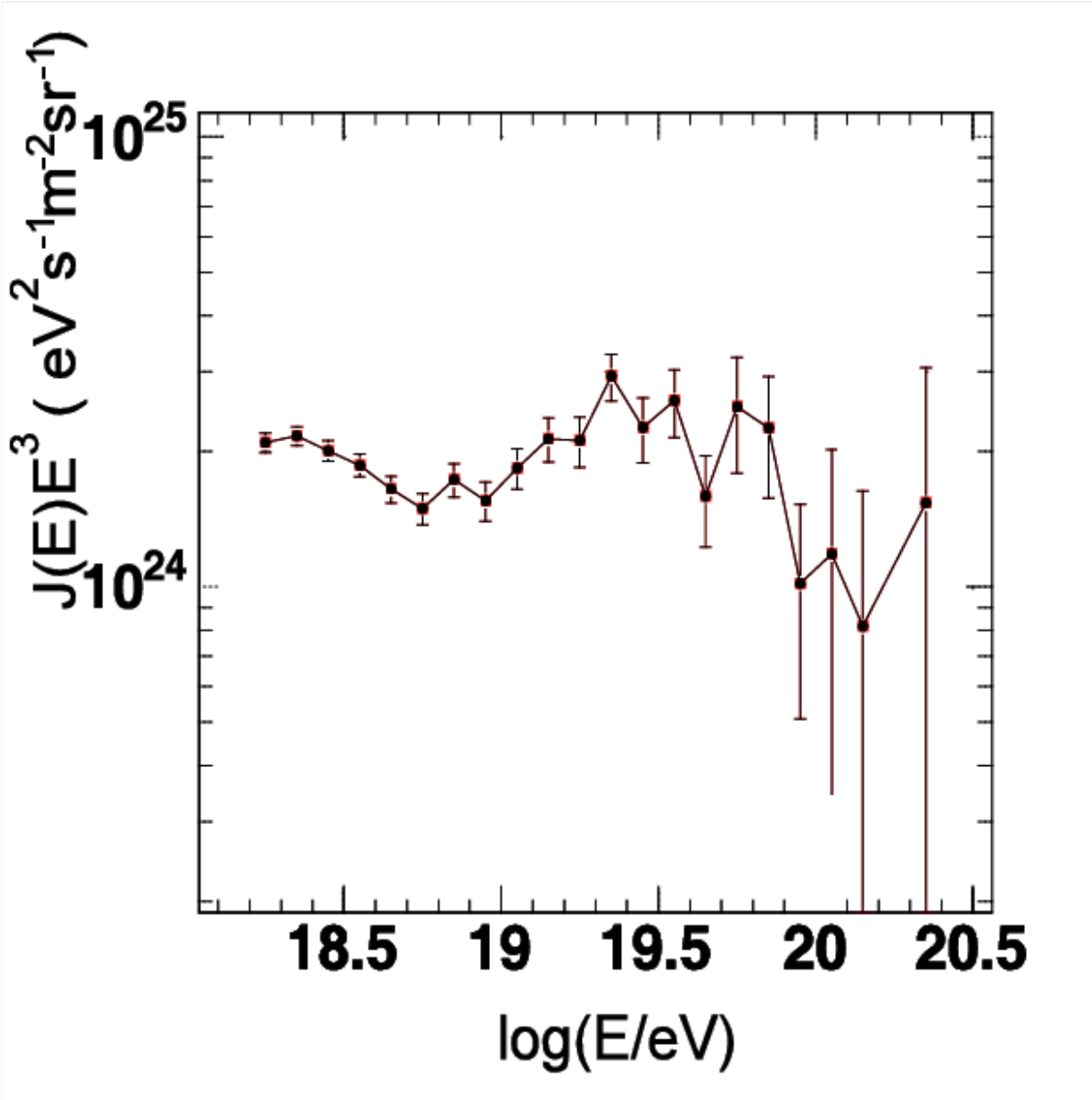

Figure 7. The HiRes stereoscopic spectrum with "fully efficient" aperture and no cloud cuts.

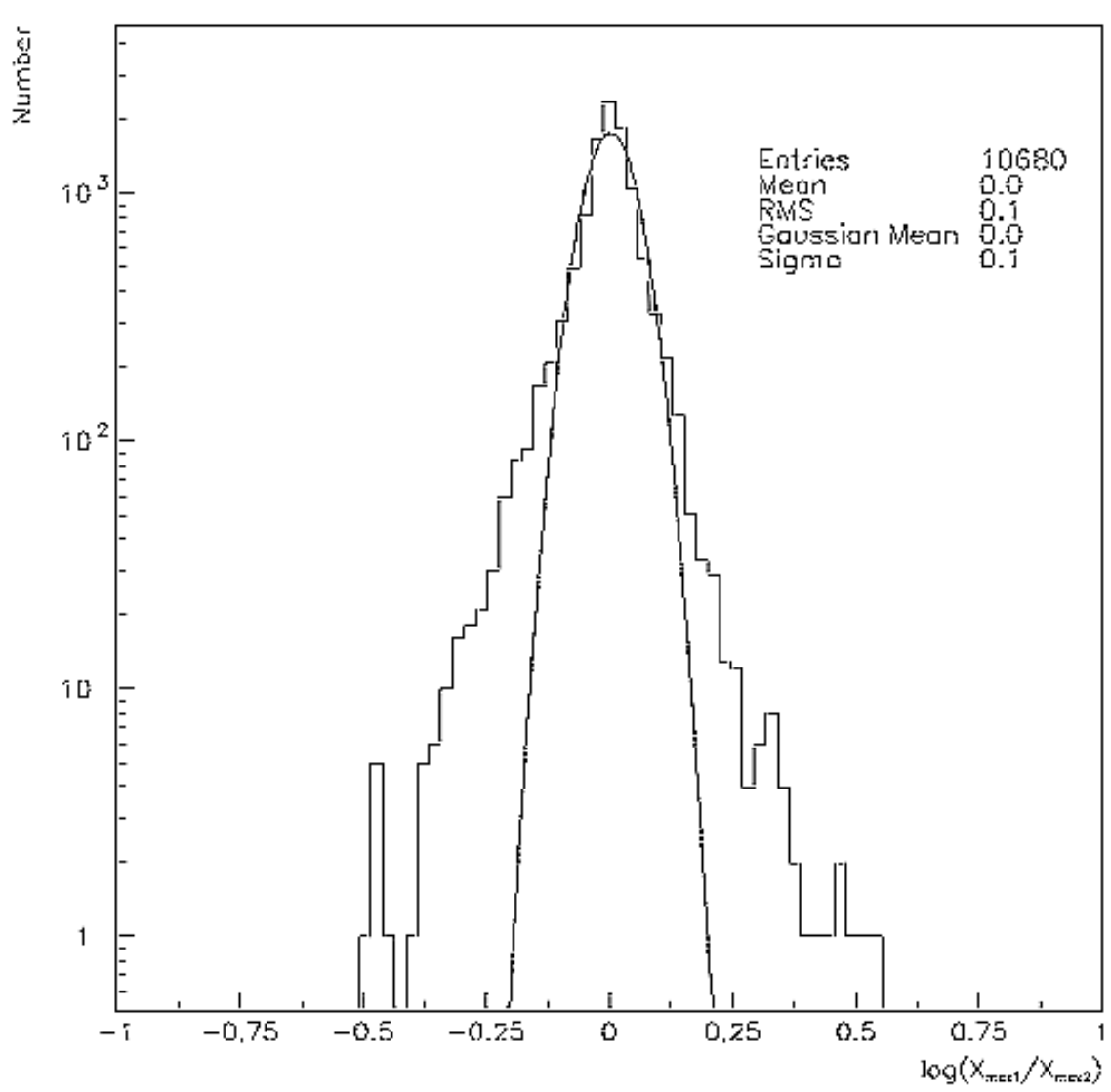

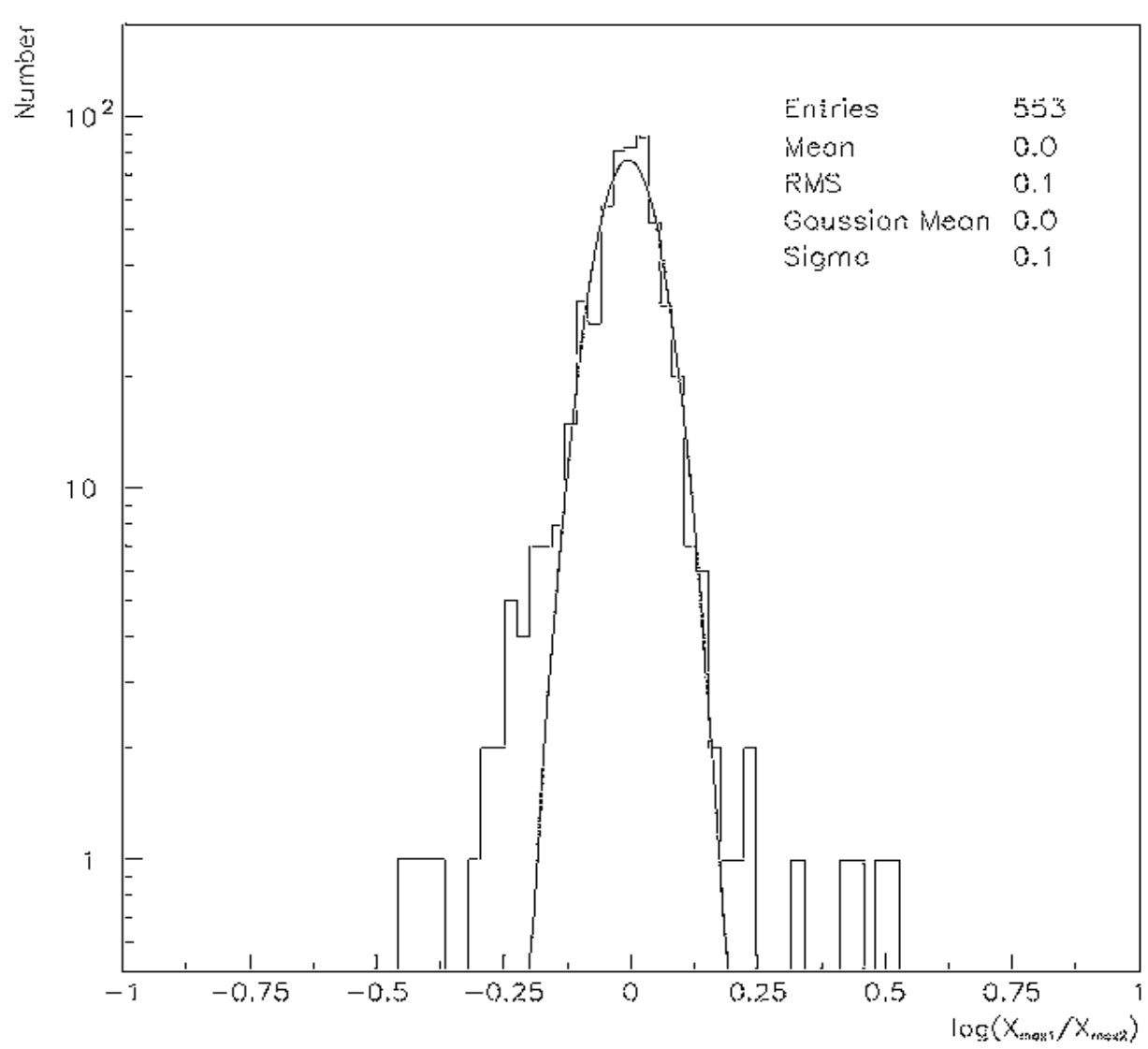

Figure 8. Pull distributions for Xmax. Log10(Xmax1/Xmax2) for simulated data (left); same variable, but for actual data (right).

**The $X_{max}$ Method for Determining Composition**

The atmosphere serves as an absorber for the energy flow of an EAS. The slope of the distribution of first interactions depends on the mass of the primary particle. At the highest energies, protons have an interaction length of near 70 gm/cm$^2$ while Iron nuclei would have an interaction length of near 35 gm/cm$^2$. While these numbers are small compared to the typical extent of an EAS (1200-1400

gm/cm$^2$), the distribution of first interactions is reflected in the distribution of depth of shower maximum $X_{max}$ in a way that amplifies this small difference.

Simulations show that proton primaries with energies near $10^{19}$ eV produce a distribution of $X_{max}$ with a mean near 700 gm/cm$^2$, while Iron primaries of the same total energy will have an Xmax mean of 80 to 100 gm/cm$^2$ shallower. Because an Iron nucleus produces an EAS which is basically a superposition of 56 lower energy showers, the fluctuations of $X_{max}$ around the mean for Iron is smaller than for protons, with Iron fluctuations having a $\sigma$ of near 30 gm/cm$^2$ and protons 70 gm/cm$^2$. Intermediate mass nuclei will lie between these numbers. While the absolute value of the position of the mean $X_{max}$ is model dependent (typical variations are 20–40 gm/cm$^2$ for protons and 20 gm/cm$^2$ for Iron at a fixed energy), the separation between the predicted mean Xmax's is much less sensitive to the model. This is also true for the difference in the fluctuations about the mean.

A related approach is to look at the so-called elongation rate of the shower maximum. Given a hadronic model and a primary particle type, the mean $X_{max}$ of an EAS will deepen linearly with the logarithm of the primary energy. This elongation is seen even in the simplest toy models of an EAS. The slope of this line, per energy decade, is called the elongation rate and is typically 50-60 gm/cm$^2$/decade independent of hadronic models and particle type [48]. A change in primary composition from heavy to light over some energy interval would produce an elongation rate larger than the 50-60 gm/cm$^2$ for an unchanging composition, while a change from light to heavy would produce a smaller or even negative elongation rate. Because of this relative insensitivity to hadronic models and to systematic errors affecting absolute $X_{max}$ determination, the measurement of the elongation rate has been the first approach to determining composition [46]. In many models, we anticipate a transition from a galactic to an extragactic cosmic ray flux somewhere between $10^{17}$ and $10^{19}$ eV [47]. This transition may be accompanied by a change from a relatively heavy composition to a relatively light one. Thus observation of a change in the elongation rate as a function of energy would be of great interest. If a change in the elongation rate can be correlated to changes in the fluctuation around the mean $X_{max}$, a convincing case can be made for a change in the cosmic ray composition in a given energy range. The precision in Xmax resolution attained by HiRes and expected for TA and Auger (~30gm/cm$^2$) is sufficient to achieve this goal.

**HiRes $X_{max}$ Measurements**

The HiRes Stereo data [48] has an energy threshold for doing physics near $10^{18}$ eV. This threshold is largely determined by the limited elevation angle of the HiRes mirrors (3-17 degrees for HiRes I and 3-31 degrees for HiRes II). Lower energy EAS develop higher in the atmosphere and a strong bias towards more penetrating EAS thus develops. A previous hybrid air-fluorescence ground-array experiment, the HiRes-Mia prototype [52], was set up to study the energy region from $10^{17}$ to $10^{18}$ eV. The MIA detector consisted of 16 muon detector patches formed with 64 scintillation counters each, covering 370x370 m with an active area of over 2500m$^2$. The patches were buried 3 m under the Earth's surface. EAS muon arrival times were recorded as well as the number of counters triggered. The threshold energy for detecting a muon was 850 MeV. The hybrid aperture for this experiment becomes fully efficient above 4 x $10^{17}$ eV corresponding to a value of 5.2km$^2$str. The MIA counters determine the muon density via the pattern of hit counters observed in the shower. An estimate of the muon density at 600m from the core, $\rho_\mu(600)$ is then determined by a fit. For these $X_{max}$ measurements, 14 HiRes I mirrors were configured in an inverted pyramid shape centered on the MIA ground array and extending from 3 deg to 70 degrees in elevation. This gave excellent $X_{max}$ acceptance over this energy range. The geometrical reconstruction of hybrid events was accomplished using monocular timing from HiRes I and MIA counter timing of the shower front arrival time. This additional timing constraint from the ground array substantially improves the precision of geometrical reconstruction over a simple monocular approach. The geometrical reconstruction of these hybrid events leads to a precision similar to HiRes stereo reconstruction. Thus in what follows, one can combine the HiRes-MIA prototype data with the HiRes stereo data. In so doing, an energy range from $10^{17}$ to near $10^{20}$ eV is covered with minimal biases in $X_{max}$.

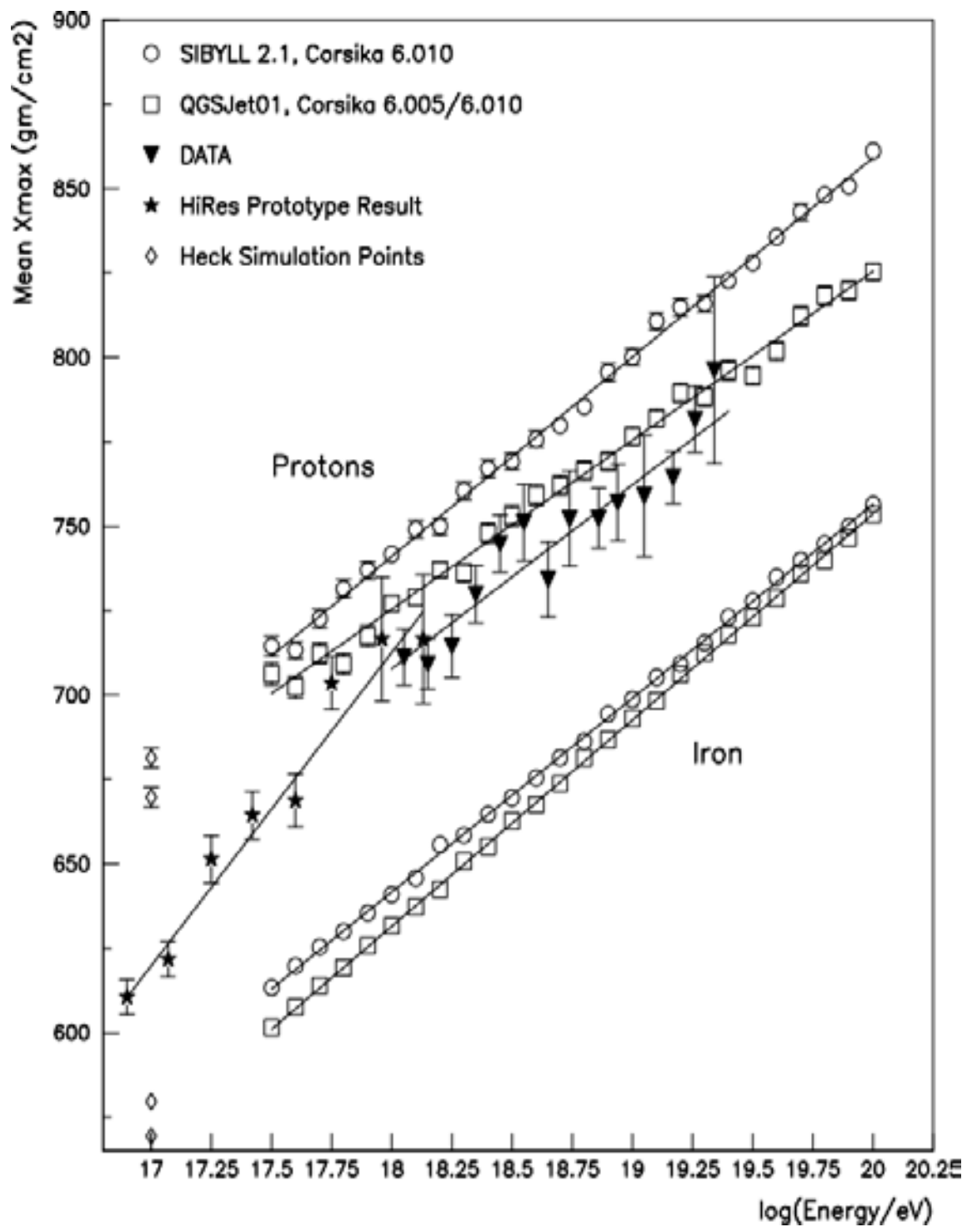

Figure 9. Elongation rate result: Theoretical predictions for protons and Iron flux and HiRes and HiRes-MIA data.

Since a pure proton and a pure Iron composition are the expected extremes of the makeup of the cosmic ray flux this is a natural place to start. To take account of hadronic model dependence, several simulations- QGSJet and SIBBYL can be compared. Proton and Iron EAS generated using these simulators are fed to the HiRes detector simulation and, if it is determined that the air-fluorescence detector triggered, "fake" data is written out in the same format as real data and analyzed using the standard reconstruction routines. In the case of HiRes-MIA prototype data, the MIA array is also simulated. Comparing reconstructed $X_{max}$ to input $X_{max}$ allows the study of the $X_{max}$ resolution and any possible biases. Predictions for the elongation rate for protons and Iron nuclei as well as their fluctuations are also generated. These can then be compared to real data.

Figures 4 and 5 show the expected $X_{max}$ and energy resolution for the Stereo HiRes data. Since data and Monte Carlo will be compared in order to extract the cosmic ray composition, it is vital that the Monte Carlo accurately reproduce the detector and reconstruction resolution. Stereo data is unique in that there are events that have $X_{max}$ independently measured by HiRes I and HiRes II. These distributions of the differences in $X_{max}$ and Energy can be used to confirm estimates of Monte Carlo resolution. Figure 8 shows the pull, defined as the difference divided by the average, for Xmax and energy, respectively, after all cuts. The nearly Gaussian shape of the pull and the nearly identical pull distribution for Monte Carlo and data show that the Monte Carlo resolution represents the real detector resolution well. For the HiRes-MIA data, no such internal crosschecks on the resolution are possible since only monocular data is available. However, the air-fluorescence part of the simulation is very similar to that used for the stereo HiRes data.

While different hadronic models produce a 10 to 30 gm/cm$^2$ shift in absolute position of the mean Xmax, the separation between the composition extremes is similar for all models. The detector Monte Carlo program and the proton and Fe profile libraries are also used to check that there is no significant Xmax bias for the energy interval of interest. Above $10^{18}$ eV and after all cuts, reconstructed Monte Carlo data leads to no significant differences in the elongation rate from the input value for either protons or Fe nuclei. Similar results are found for the HiRes prototype detector simulation.

Figure 9 shows the real data elongation rate after data quality cuts (identical to the Monte Carlo data cuts). Above $10^{18}$eV the measured elongation rate (ER) is 54.5 ± 6.5 gm/cm$^2$/decade (statistical errors only). The systematic errors on the elongation rate are estimated to be 6 gm/cm$^2$/decade. This is to be compared to 50 and 61 gm/cm$^2$ for QGSJet protons and iron nuclei, respectively and 57 and 59 gm/cm$^2$ for SIBYLL protons and iron nuclei. The data above $10^{18}$ eV are thus quite consistent with an unchanging composition. The HiRes-MIA prototype data at lower energies gives an elongation rate of 93.0 ± 8.4 gm/cm$^2$/decade, strongly suggesting a change in composition below $10^{18}$eV in the direction of an increasingly heavy composition near $10^{17}$ eV.

Figure 9 also shows that the absolute value of the mean $X_{max}$ at any energy above $10^{18}$ eV is in much better agreement with a light, mainly protonic composition than with a heavy, iron-dominated one. The systematic errors on the absolute mean $X_{max}$ measurement are 18.7 gm/cm$^2$, while the differences between QGSJet01, QJSJetII and SIBYLL give an indication of the effect of different theoretical assumptions in the hadronic models. A more quantitative statement about composition can be made by comparing the distribution of measured $X_{max}$ over some energy range to the expectation for pure protons or Iron. Fig 10 shows the Xmax distribution for energies > $10^{18}$ eV. Lower energy experiments such as Fly's Eye and HiRes/MIA have larger intrinsic $X_{max}$ resolutions that the HiRes Stereo result and hence direct comparison is difficult and awaits new experiments. The fluctuations in the data for energies greater than $10^{18}$ eV are much larger than what is expected for pure Iron and a two-component fit to the distribution yields 80% protons for QGSJet and 60% for SIBYLL.

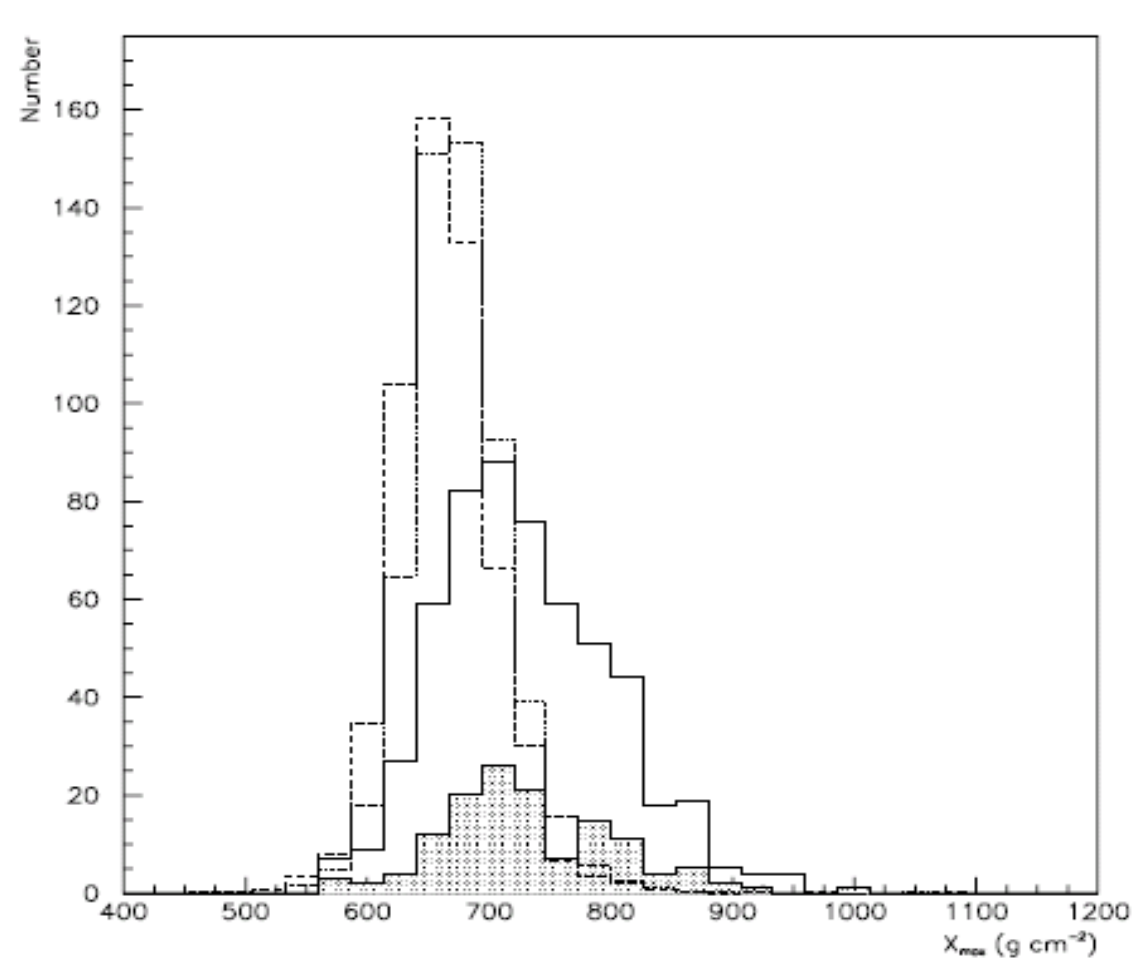

Figure 10. Xmax distribution from the HiRes stereo experiment for events with energies > $10^{18}$ eV. Solid line is the data. Dotted and dash-dotted lines are predictions for pure Fe for two hadronic models normalized to the same area as the data. Shaded histogram corresponds to events where an average atmospheric corrections (as opposed to hourly) was made. Data is incompatible with a heavy composition.

**Comparison with Fly's Eye results**

Because a change in the CR composition below $10^{18}$ eV may be an important indicator of a transition between a galactic and extragalactic cosmic ray flux (or at the least a change in the nature of the CR source itself), it is of great importance to confirm this result. While a new experiment with an energy range covering from below $10^{17}$ eV to $10^{20}$ eV would be the ideal and has been proposed [49], the previous Fly's Eye experiment [46] stereo measurements of the elongation rate and fluctuations in the energy range $10^{17}$ to $10^{19}$ eV are of interest, even though they have significantly worse Xmax resolution (45 gm/cm$^2$). There are quantitative differences between the Fly's Eye and the HiRes and HiRes-MIA result. To see if these differences are significant, the quoted systematic errors on the absolute Xmax measurement for both experiments need to be taken into account [50]. Since shower profiles measured above $10^{18}$ eV had the best resolution in energy and Xmax a comparison of Fly's Eye and HiRes stereo mean Xmax in this energy range gives the best indication of systematic shifts between the two experiments. The data comparison indicates a systematic shift of 14 gm/cm$^2$, well within the quoted systematic errors. The entire Fly's Eye data can then be shifted upward by this amount. Figure 11 presents the result, which now shows generally good agreement between HiRes and Fly's Eye throughout the entire energy range. The combined HiRes, HiRes-MIA prototype and Fly's Eye data clearly show a transition from a heavy to a light composition somewhere between $10^{17}$ and $10^{19}$ eV. The HiRes data, with better Xmax resolution, prefers a more abrupt transition. The Xmax distribution fluctuations are also quite consistent between Fly's Eye, HiRes-MIA and HiRes and show evidence for broadening with higher energy, consistent with the trend towards a lighter composition.

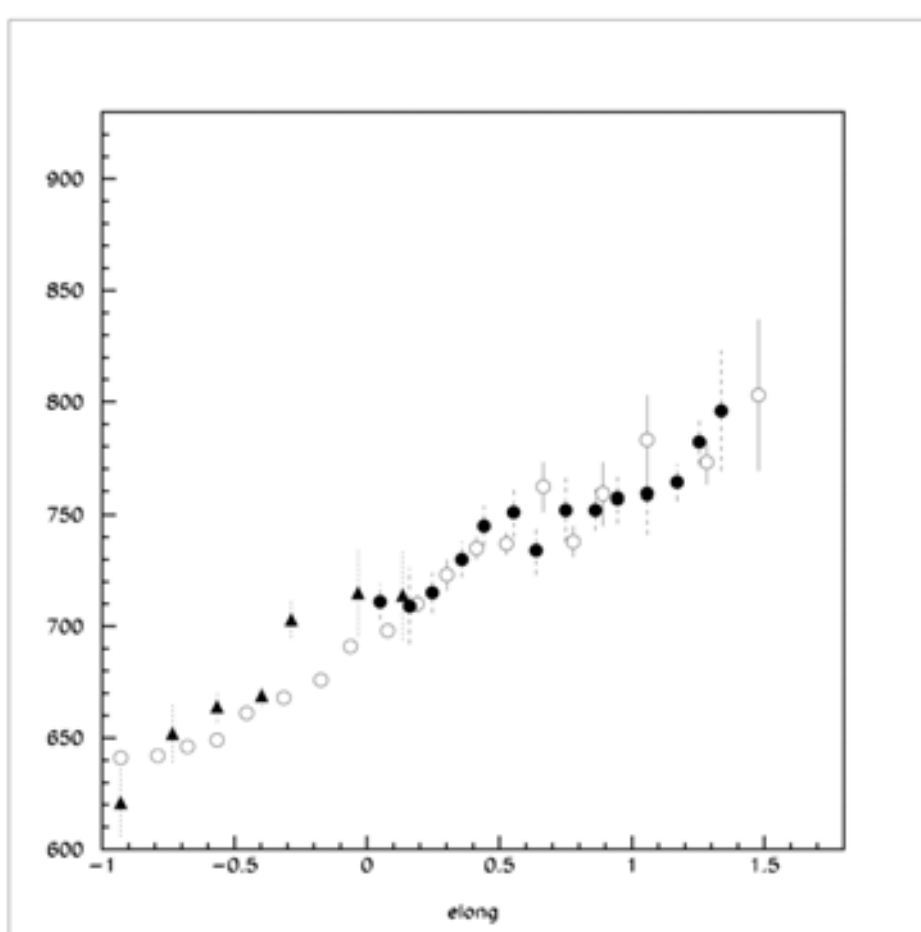

Figure 11. Comparison of Fly's Eye, HiRes-MIA, and HiRes. Open circles – Fly's Eye data shifted up by 14 gm/cm$^2$; Filled triangles – Hires Mia data; Filled circles – HiRes data.

**Composition from Ground Array Information**

Surface arrays can measure the muon content of EAS. The muon density at a given distance from the core has sensitivity to the nature of the primary particle. Simulations indicate that the muon content of Iron initiated showers, at sea level and with a muon threshold energy of 1 GeV is about 1.4 times greater than for protons for energies near $10^{17}$ eV. The muon density increases linearly with log(E) and the slope of this relation can play the same role as the elongation rate for Xmax, vis:

$$\beta = \mathrm{dlog}(\langle \rho_{\mu}\,(600)\rangle)/\mathrm{dlogE}.$$

A shower initiated by a nucleus of mass number A and energy E is a superposition of A subshowers each with energy E/A and thus $\rho_\mu$ (600) is proportional to $A(E/A)^{\beta 0}$, where the value of $\beta_0$ is dependent on the hadronic interaction model, but in practice is largely independent of A. Hence a deviation of β from what is expected for a pure composition implies a changing composition, i.e. dlogA/dlogE = $\beta$-$\beta_0$/(1-$\beta_0$).

Two recent experiments, AGASA and the HiRes-MIA prototype have reported results on the muon content of EAS above $10^{17}$ eV [51] [52]. We describe these in turn.

The unique feature of the HiRes-MIA experiment is its hybrid nature. The energy of the primary particle is measured by the air-fluorescence technique, while the underground muon detector array samples the muon density at ~ 1 km from the shower core. $X_{max}$ information is also available from the HiRes detector. The results are shown in figure 12, with β = 0.73 +/-0.03+/-0.02/decade. Also shown in the figure are results from simulations using the Corsika package and the QGSJet and SIBYLL hadronic interaction models. The QGSJet model after reconstruction and all cuts gives an β = 0.83+/-.01/decade for both protons and iron. SIBYLL predicts significantly fewer muons for both protons and iron showers. Data and simulations are clearly inconsistent with each other in the muon content index beta. While this can be interpreted as supporting a change towards a lighter composition, consistent with the $X_{max}$ measurements described above, there is a problem with the absolute density of muons at 600m. The data shows values of $\rho_\mu$ (600) at lower energies, which are larger than what is predicted for Fe.

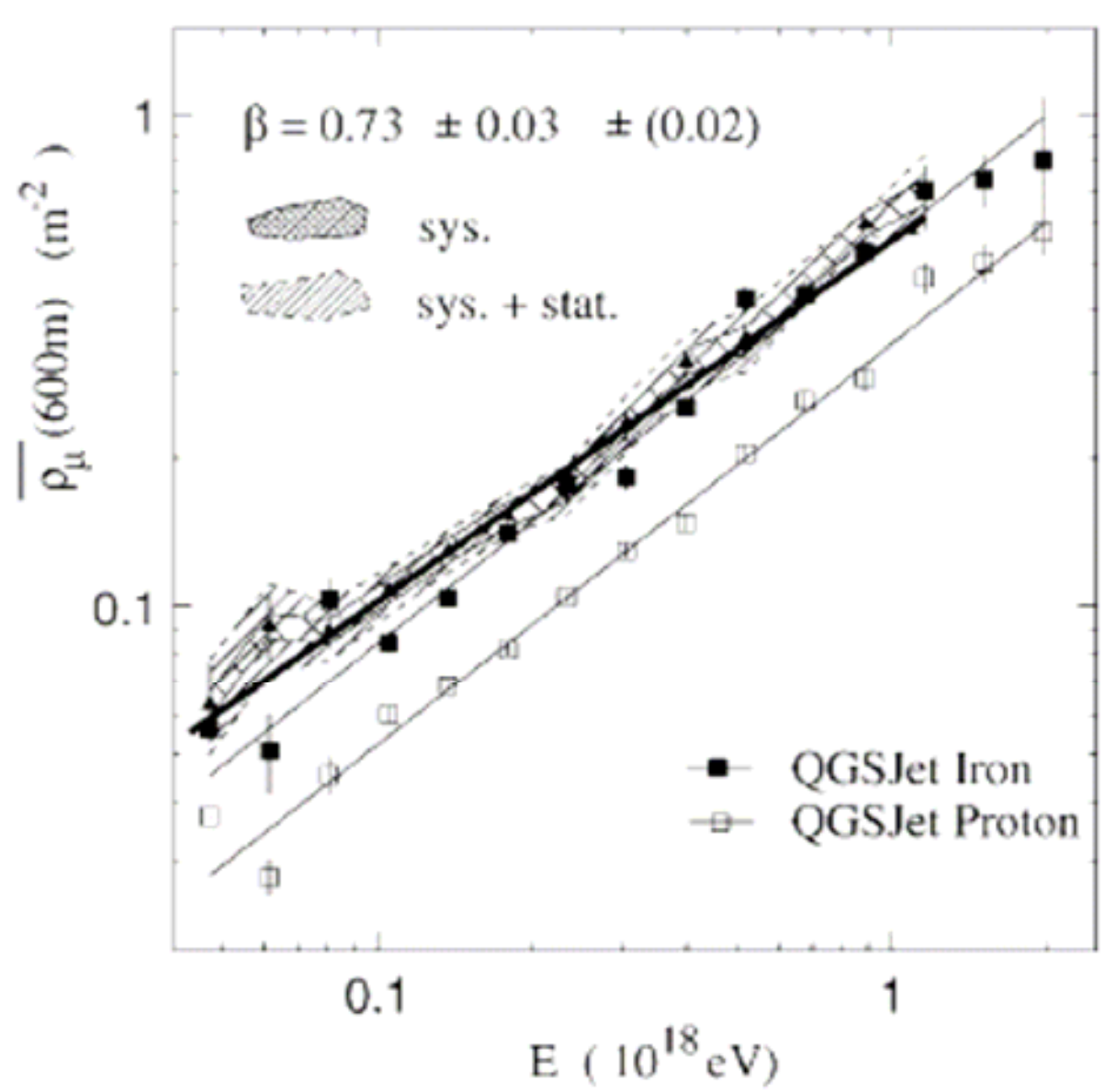

Figure 12. HiRes-MIA result on $\rho_\mu$ (600), the muon density at 600 meters from shower core, as a function of energy.

A similar problem appears with the combined Akeno and AGASA muon data [53]. The Akeno 1 $km^2$ ground array was instrumented with 156 one $m^2$ scintillation counters and eight 25$m^2$ surface muon detectors covered with 2 m of concrete. This experiment had an effective energy range between $10^{165.}$ eV to $10^{18.5}$ eV. The Akeno experiment relied on the total shower size $N_e$ at observation level to estimate the primary energy rather than ρ(600). In the AGASA experiment the relation between muon density $\rho_\mu$ (600) and electron density ρ(600) is shown in Fig 13, together with predictions using the QGSJet and SYBILL models. This figure shows that AGASA data is in fact marginally consistent with a transition between heavy and light composition between ρ(600) of 1 and 100 $m^{-2}$ (corresponding to an energy range of 2 x $10^{17}$ to 2 x $10^{19}$ eV), since the experimental slope is flatter than the predicted one for a constant composition. The difficulty is that at energies below $10^{17}$ eV, the predicted value of $\rho_\mu$ (600) for Iron with either hadronic model will be smaller than the extrapolation from measured data. The lower energy Akeno data can be used to check if that is in fact the case.

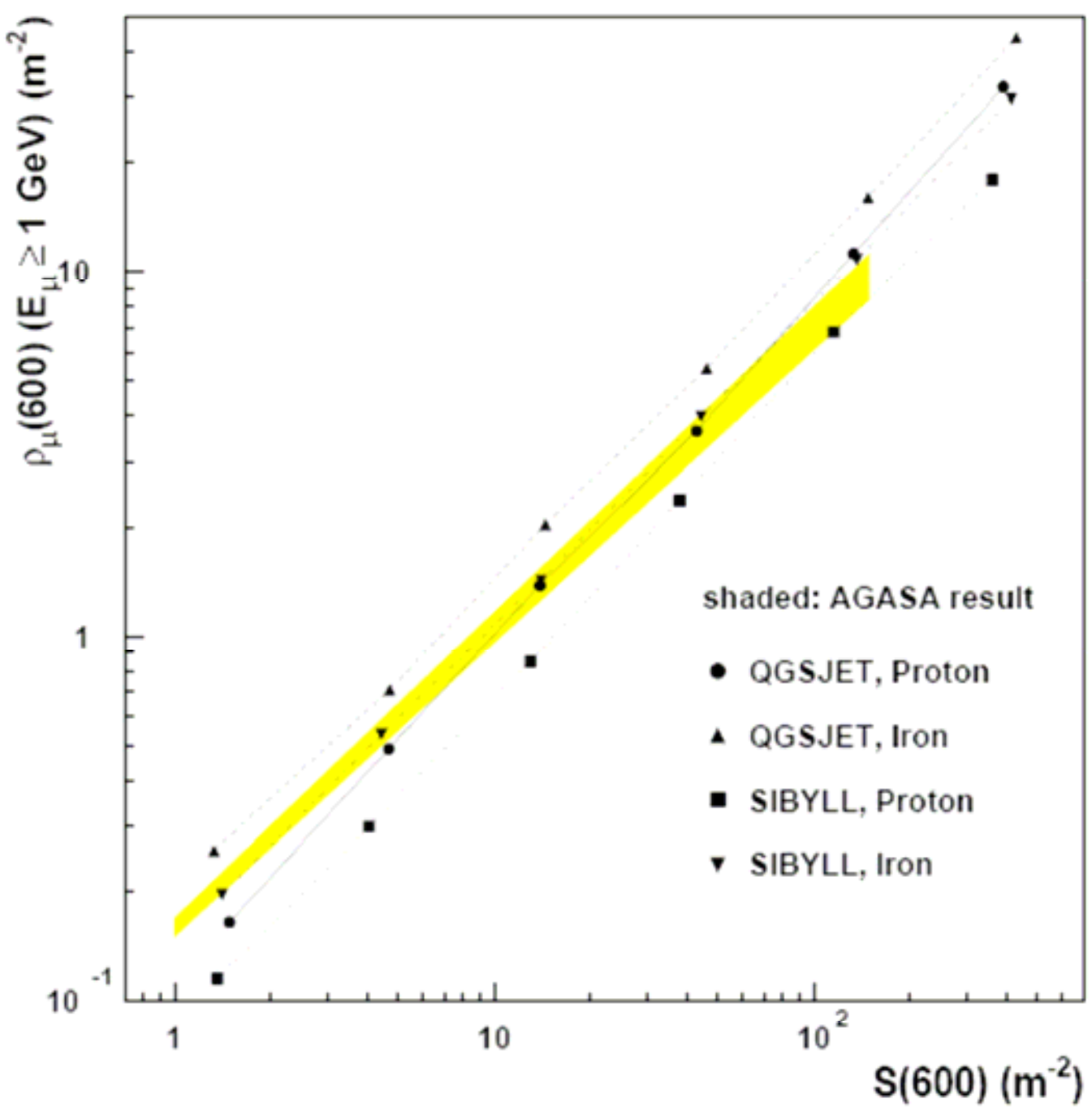

Figure 13. AGASA data on muon density as function of energy estimator S(600) and comparison with predictions.

However, since the Akeno experiment did not use ρ(600) as the energy estimator, but rather the total shower size at ground level, a systematic difference in energy scales between the low energy Akeno and the higher energy AGASA data is possible. Barring such a shift, the ρ(600) and $N_e$ energy estimators can be cross-correlated and the muon density over a wider range of energies can be explored. The result is that the measured Akeno composition becomes heavier than that predicted for iron below $10^{17}$ eV [54].

Recently, a new hadronic generator, EPOS, has been added into the Corsika framework [36]. EPOS is based on a parton/string model and is unique in that it incorporates an explicit treatment of baryon-antibaryon production. The baryon part of the simulation is consistent with RHIC data and simulates the observed transverse momentum broadening, parton saturation and screening and other effects [55]. It is also consistent with other accelerator measurements. This model leads to a significant increase in the predicted muon density, with a 40% increase near $10^{17}$ eV. There is very little change in the $X_{max}$ predictions, however. The effect is due to increasing the baryon-antibaryon content of the EAS and hence decreasing the energy flow to pions and kaons. A rapid increase in the nuleus-antinucleus production was first observed in air shower experiments studying delayed hadronic signals near air-shower cores[56]. EPOS predictions are in excellent agreement with the HiRes-MIA $\rho_\mu(600)$ data and leads to a self-consistent conclusion about the cosmic ray composition using both $X_{max}$ and $\rho_\mu(600)$ data. Work is in progress to compare with AGASA and Auger measurements.

**The p-Air Inelastic Total Cross-Section**

Several techniques have been proposed and used to measure the total inelastic proton-Air cross-section using ultra-high energy cosmic ray data. Results from AGASA [57], the Fly's Eye [58] experiment and the stereo HiRes experiment [59] utilize very different approaches. Here we will discuss the HiRes analysis, as it is the least model dependent.

The slope of the distribution of first interaction in the atmosphere, $\lambda_{p\text{-}air}$ is inversely proportional to $\sigma_{pair}$. This distribution, however, is not measurable by the air fluorescence or any other indirect observation technique. What can be measured with good accuracy is the statistical distribution of $X_{max}$. The mean of this distribution is used to extract information on composition. The decrement $\Lambda$, or slope of the distribution beyond the peak of the $X_{max}$ distribution, is sensitive to the value of $\lambda_{p\text{-}air}$ with $\Lambda = K\lambda_{p\text{-}air}$ where the value of K is model dependent and ranges from 1.2 to 1.6 [60].

A less model dependent approach was recently developed by the HiRes collaboration. This relies on the fact that the most of the EAS shower development is governed by hadronic interactions and decays at much lower energies, than the energy of the primary particle where most models are in good agreement. Near shower maximum, for example, the mean electron energies are order of 1 GeV to 100 MeV (the critical energy in air). Separating the effects of the first interaction from the subsequent EAS development can lead to more reliable results.

The observable $X_{max}$ distribution can be considered to be a convolution of two distributions. The first is the distribution of the depth of the first interaction, which can be approximated by an exponential function

$$N_1(x) = \exp(-x/\lambda_{p\text{-}air}).$$

The second is the distribution in depth variable $X' = X_{max} - X_1$ where $X_1$ is the depth of the first interaction. The distribution in X' can be calculated in Monte Carlo simulations and can be approximated by a power-exponent function of three parameters. These parameters can be studied as a function of energy in Monte Carlo simulation and thus the function of X' becomes a 'known" function. As expected, the three parameters are largely independent of different hadronic model assumptions.

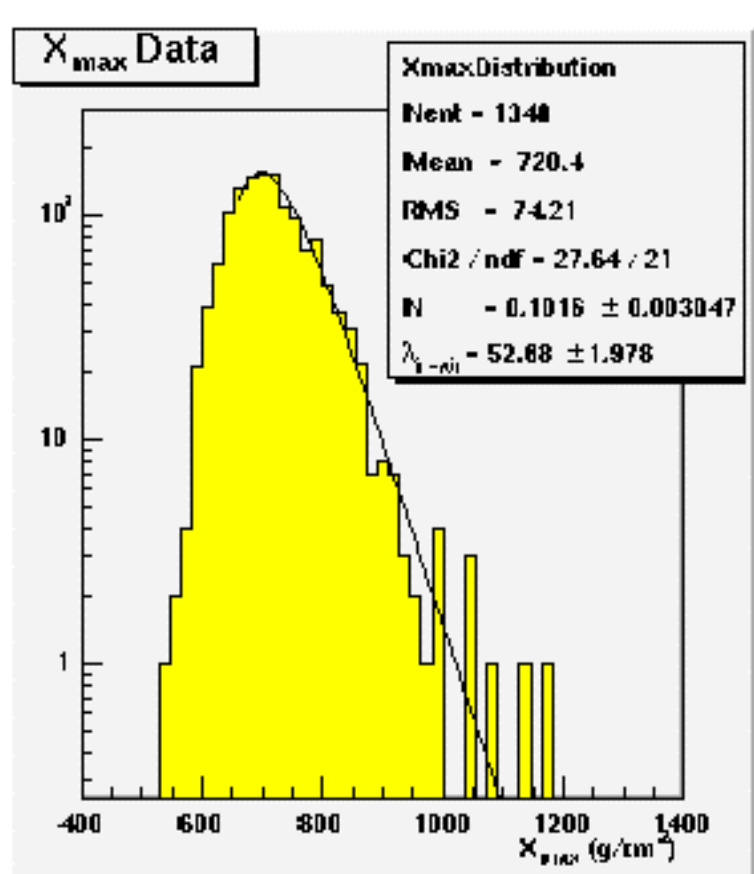

Figure 14. Distribution of Xmax and fit using deconvolution method.

The $X_{max}$ distribution is then a convolution of known functions with one unknown parameter, $\lambda_{p\text{-}air}$ Simulations show that the input value of $\lambda_{p\text{-}air}$ is correctly found, within statistical errors, by this method, for both QGSJet and SIBBYL models. Figure 14 shows the fit to the data and the value of $\lambda_{p\text{-}air}$ extracted from deconvolution.

The proton-air inelastic cross section found by this method at 3 x $10^{18}$ eV is shown in figure 15 together with measurements by Akeno and the Fly's Eye results and lower energy data.

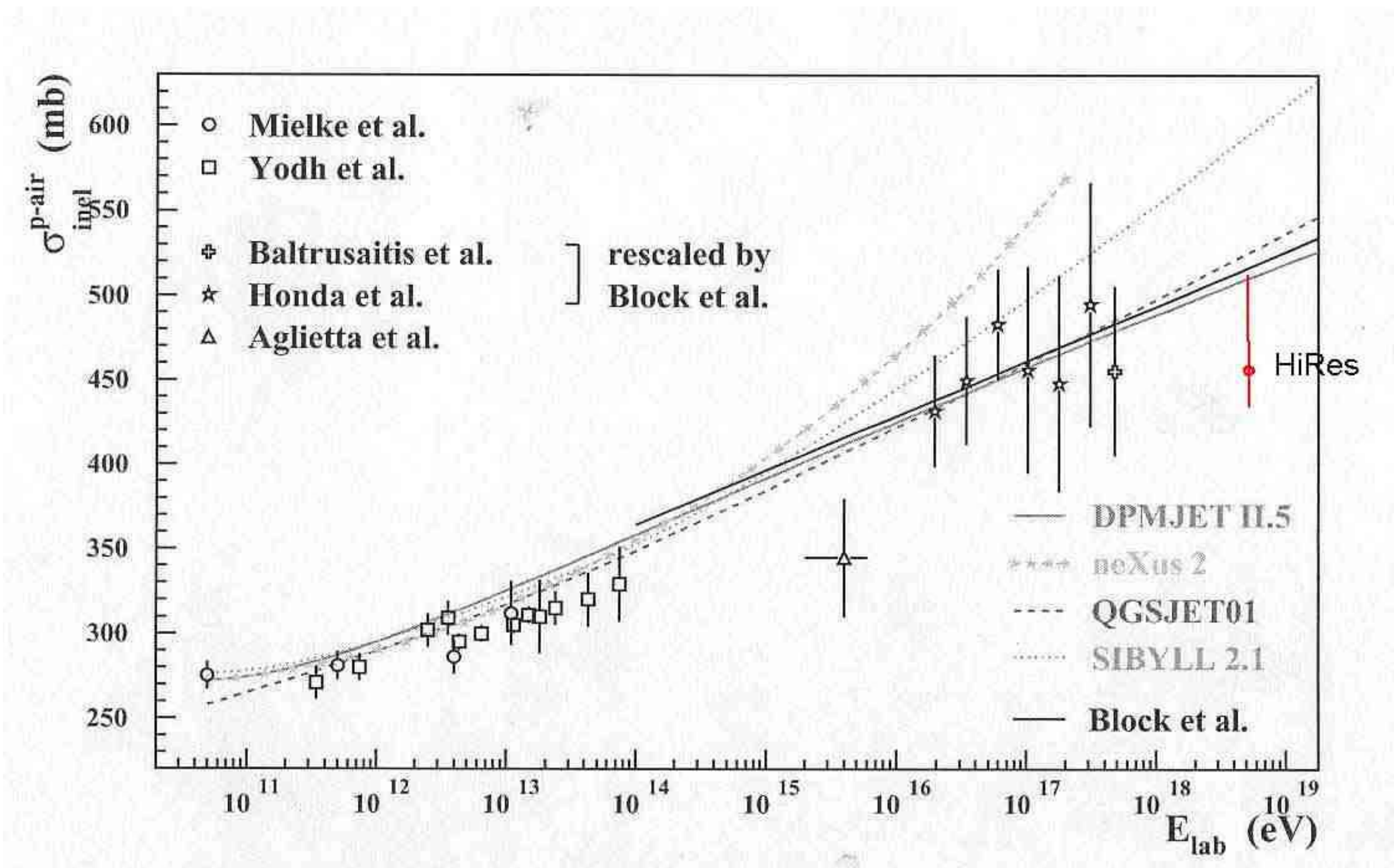

Figure 15. Inelastic cross-section measurements and predictions from a variety of hadronic models. HiRes data point includes systematic errors from possible 5% gamma-ray contamination.

**Small Scale Anisotropy**

One of the outstanding questions of cosmic ray physics is, "What are the sources of these particles?" For galactic cosmic rays a growing consensus maintains that the sources are supernova remnants [61]. Recent Hess TeV gamma ray results showing flatter than expected spectra from many extended galactic objects are consistent with this expectation [62]. The remnants of O and B stars, in "superbubbles," may play a particularly strong role in accelerating galactic cosmic rays [63]. The supermassive black hole at the center of the Milky Way galaxy is said to be only about $10^6$ solar masses, and has low luminosity, so it is not thought to contribute to the cosmic ray flux in a significant way [64]. Supermassive black holes in other galaxies, however, are one of the main candidates for accelerating extragalactic UHE cosmic rays [65]. If this is true then cosmic rays of sufficiently high energy may propagate in almost straight-line trajectories through the extragalactic and galactic magnetic fields from these sources and point back to them. Every cosmic ray experiment carries out a search for small-scale anisotropy (e.g., point sources, possibly blurred by magnetic bending).

The AGASA experiment found a large autocorrelation signal among their events at energies greater than 4 x $10^{19}$ eV. They also found seven clusters of events, six doublets and a triplet, where they would expect between one and two from random coincidences given the statistics of their data. Several models of cosmic ray sources were devised to explain these results, one of which consisted of a count of the minimum number of sources needed to produce the clustering result (it is about 200).

A similar autocorrelation searched was performed for the HiRes data and no significant signal was seen [66]. To understand this discrepancy better, the HiRes collaboration carried out a study [67] whose ansatz was that the seven AGASA clusters represent the seven brightest sources of cosmic rays. If this were true then HiRes would be likely to observe events correlated with those of the AGASA clusters. This study used the HiRes stereo data with selection criteria similar to those of AGASA. Because of the discrepancies in the normalization of the cosmic ray spectrum between the two experiments, an exact match is not possible because there is no linear transformation (such as an energy scale shift), which can bring the numbers of events observed by AGASA and HiRes into agreement as a function of energy. The AGASA events have an angular resolution of about 2°, while the HiRes stereo event resolution is about 0.6°. The result of this study was that no HiRes events of energy above 4 x $10^{19}$ eV were correlated with the AGASA clusters. However, an event was found, with an energy of 3.7 x $10^{19}$ eV, right in the middle of the AGASA triplet. The AGASA triplet, plus this HiRes event, are now called the "quartet". The chance probability of this occurring is about $10^{-3}$. A sky map of the quartet is shown in figure 16. The colors at different locations on the sky indicate the probability of finding the center of the quartet at that location. The pointing accuracy of the HiRes stereo event dominates the determination of the most likely location, which is (RA, dec) = (169.1 +-0.6, 56.3+-0.4) degrees. It will be interesting to see if TA sees something at this location in the sky. The quartet is in Ursa Major, and is not visible from the location of the Auger experiment.

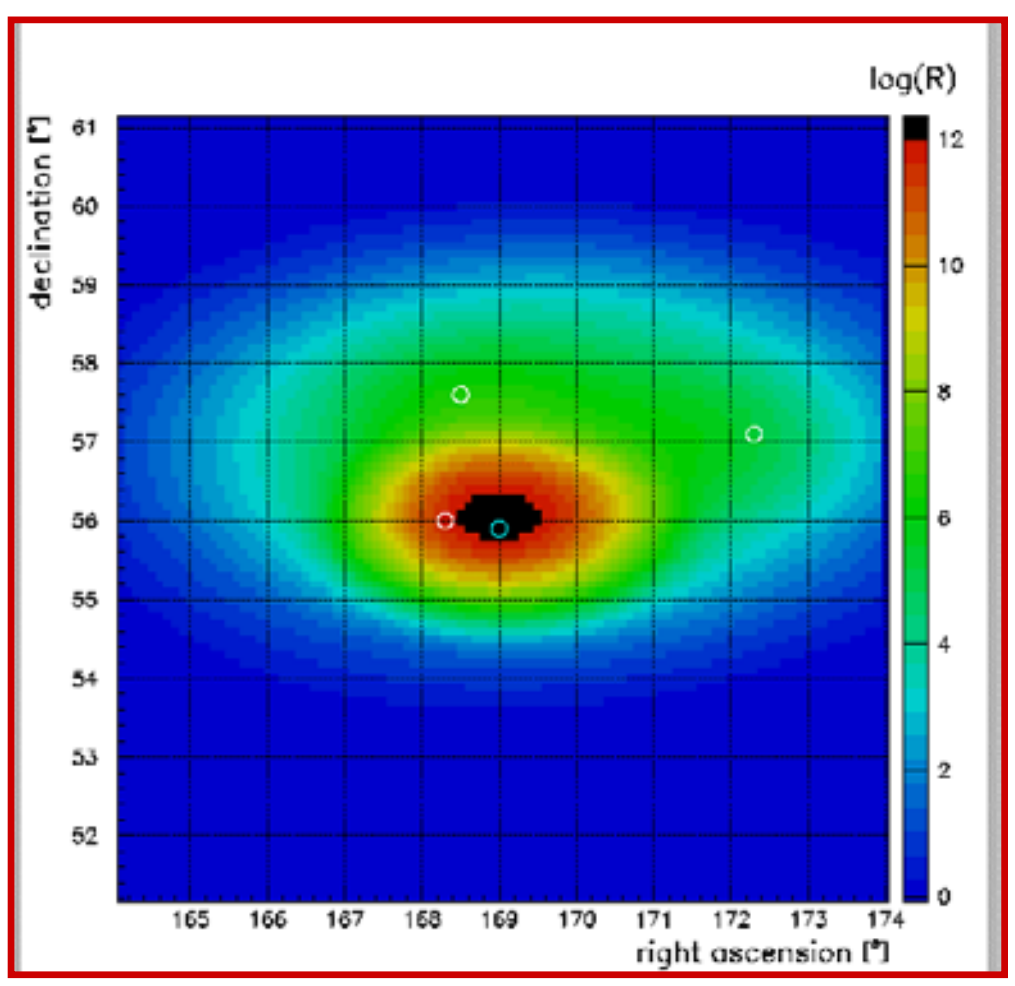

Figure 16**.** Sky map of the AGASA and HiRes events making up the "quartet." The HiRes event is at the center of the black region. The color scale represents a likelihood ratio, with black being logR=12, red 11,yellow 9 , green 6 and blue zero (see text).

**Correlations with BL Lac Objects**

I. Tkachev, P. Tinyakov, D. Gorbunov, and others, pioneered the search for correlations between cosmic ray events and BL Lac sources. BL Lac's (also called blazers) are extragalactic astronomical objects (highly variable AGN's) with their jets pointing directly toward the earth. They are good candidates for cosmic ray accelerators because of this geometry. Tkachev et al. [68], and D. Gorbunov et al. [69] first searched the Yakutsk and AGASA data above 4.0 x $10^{19}$ eV using the Veron catalog of BL Lac objects, and found correlations at the $10^{-4}$ chance probability level. They then searched HiRes stereo data above $10^{19}$ eV, assuming a uniform angular resolution of 0.8°, and found correlations at the $10^{-4}$ chance probability level [70]. In these studies they made one cut on the Veron catalog events, choosing BL Lac's brighter than magnitude 18. Such sources are described in the catalog as being better identified BL Lac's.

The HiRes collaboration repeated the correlation study taking into account the pointing resolution of each event and was able to reproduce the result described above [71]. There is an additional category of sources in the Veron catalog, BL Lac's with high polarization (so-called HP sources), which Tkachev and collaborators neglected. Upon including the HP sources correlations were found at the chance probability level of $10^{-5}$. These studies were carried out with HiRes stereo data collected through January 2004. Analysis of the remainder of the HiRes data, and studies of other properties, such as the energy spectrum of all events and correlated events, are ongoing.

There are several interesting aspects of BL Lac correlations. One is that those with Yakutsk and AGASA data are consistent with the bending of cosmic rays by galactic magnetic fields (at about 1° for protons), but the HiRes stereo events have pointing resolution about 1/3 of the expected bending angles. So, if the correlations are real, about 3% of cosmic rays must be neutral, or the galactic magnetic fields must be smaller than expected, at least in some directions. In addition, searching for correlations with BL Lac sources is mostly a northern hemisphere activity since only about 1/3 as many BL Lac sources are known in the southern hemisphere. The figure of merit for BL Lac searches is $A/R^2$, where A is the aperture and R is the angular resolution. Cosmic ray events seen in hybrid and stereo modes simultaneously have the best resolution, so future experiments should try to optimize their hybrid-stereo aperture. Further data (and future experiments) in this area will be very interesting.

**Large Scale Anisotropy**

The AGASA experiment saw an interesting anisotropy signal at larger angular scales [72]. In the Raleigh analysis of their data in a narrow energy band (about a factor of 2 wide) about $10^{18}$ eV, the AGASA experiment saw an excess of events near (but not at) the galactic center and a deficit of events near but not at the galactic anticenter. The events were integrated over 20° for this study. The narrowness of the energy band raises doubts about the reality of such an effect. These two regions have now been examined in the HiRes and Auger data. The Auger collaboration analyzed their data in a similar way to AGASA and published a non-observation of excess near the galactic center [73]. In the northern hemisphere this region of the sky is observed in summer, which reduces observation time because of short nights. HiRes does not have the sensitivity to make statements about anisotropy near the galactic center.

However, near the galactic anticenter, in Scorpius, HiRes sees a deficit of events in almost the same place as the AGASA deficit. Figure 17 shows the AGASA and HiRes sky maps in right ascension and declination. Here the HiRes data have not been limited in energy, but include the whole energy range, 10,326 events in all. However the HiRes energy histogram peaks at $10^{18}$ eV, just at the center of the AGASA energy band.

One usually thinks of anisotropy studies as searching for excesses, not deficits, in the data. One way a deficit might occur is through the galactic magnetic field integral being large enough in some directions to exclude extragalactic cosmic rays coming from those directions. This in fact occurs with some field models [74]. Again, the TA experiment will have interesting things to say about this possible anisotropy signal. The galactic anticenter appears at a zenith angle of 70° for the Auger experiment, so it is unlikely that they can observe this part of the sky with good statistics and systematics.

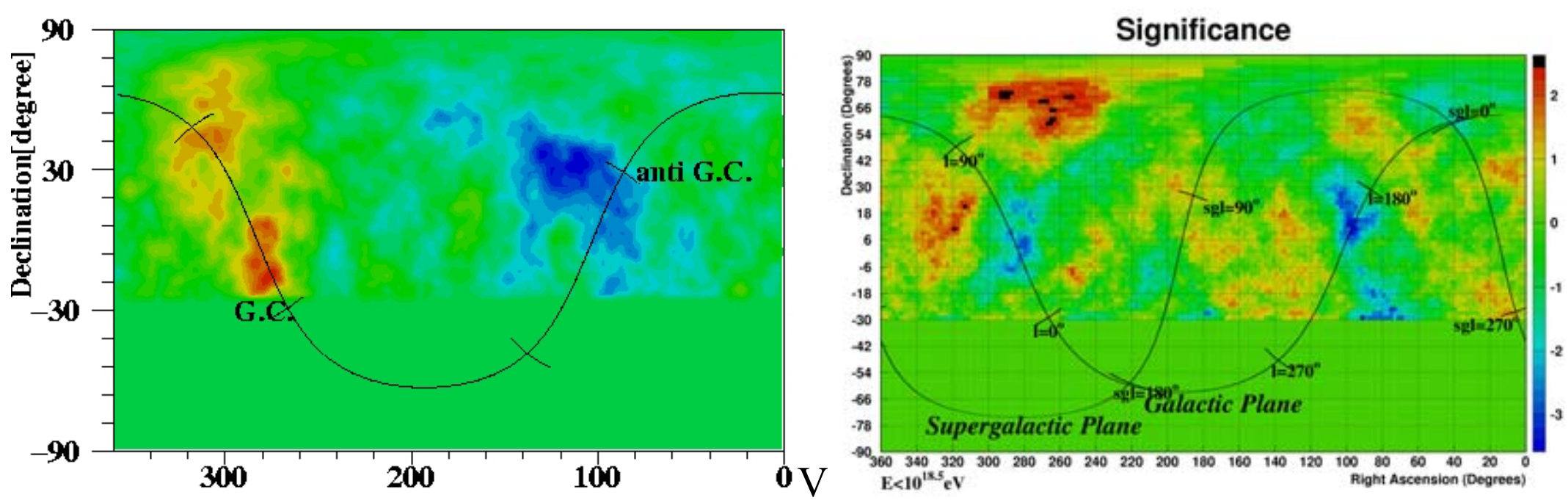

Figure 17. Significance maps of AGASA (left) and HiRes (right) showing the deficit near the galactic anticenter (RA=100°, δ=18°) in both data. Right ascension is shown on the x-axis and declination on the y-axis.

**Shower Development Profile**

While most of the hadronic interactions and decays that produce an EAS are at low energies, it is important to validate the complex simulation programs that are required to predict such profiles. The HiRes/MIA hybrid data was used to study the development profile of showers [75]. Shower profiles of well-measured events were normalized to each other in terms of maximum shower size and depth of maximum and averaged. To normalize out the Xmax fluctuation, the shower was rescaled in terms of the shower age parameter. The resultant mean shower and the predicted functional form (the Gaisser-Hillas function) are in good agreement over most of the longitudinal development.

The HiRes monocular data (from HiRes-II) has also been used to study the development profile of showers. Here a different approach was taken than in the HiRes/MIA case: a direct comparison was made between data and Monte Carlo events generated using Corsika and QGSjet. This is made possible by the excellent agreement between HiRes data and our Monte Carlo simulation. Shower age is again used for the comparison. After averaging over events, the data/MC comparison of the age distributions for all energies is shown in the left part of figure 18. The ratio of data to MC is shown in the lower part of the figure. The agreement is good.

The HiRes monocular data allow one to make these comparisons for a wider energy range than was possible with HiRes/MIA. The width of data and MC age distributions is shown in the right part of figure 18, for each half-decade in energy. The error bars in this figure are the statistical uncertainties. In addition, there are systematic uncertainties of about 0.005. A discrepancy in shower widths is apparent for the highest energy showers. The size of the discrepancy is about four times the systematic uncertainty. As there are more systematic studies to be done, this result is still preliminary.

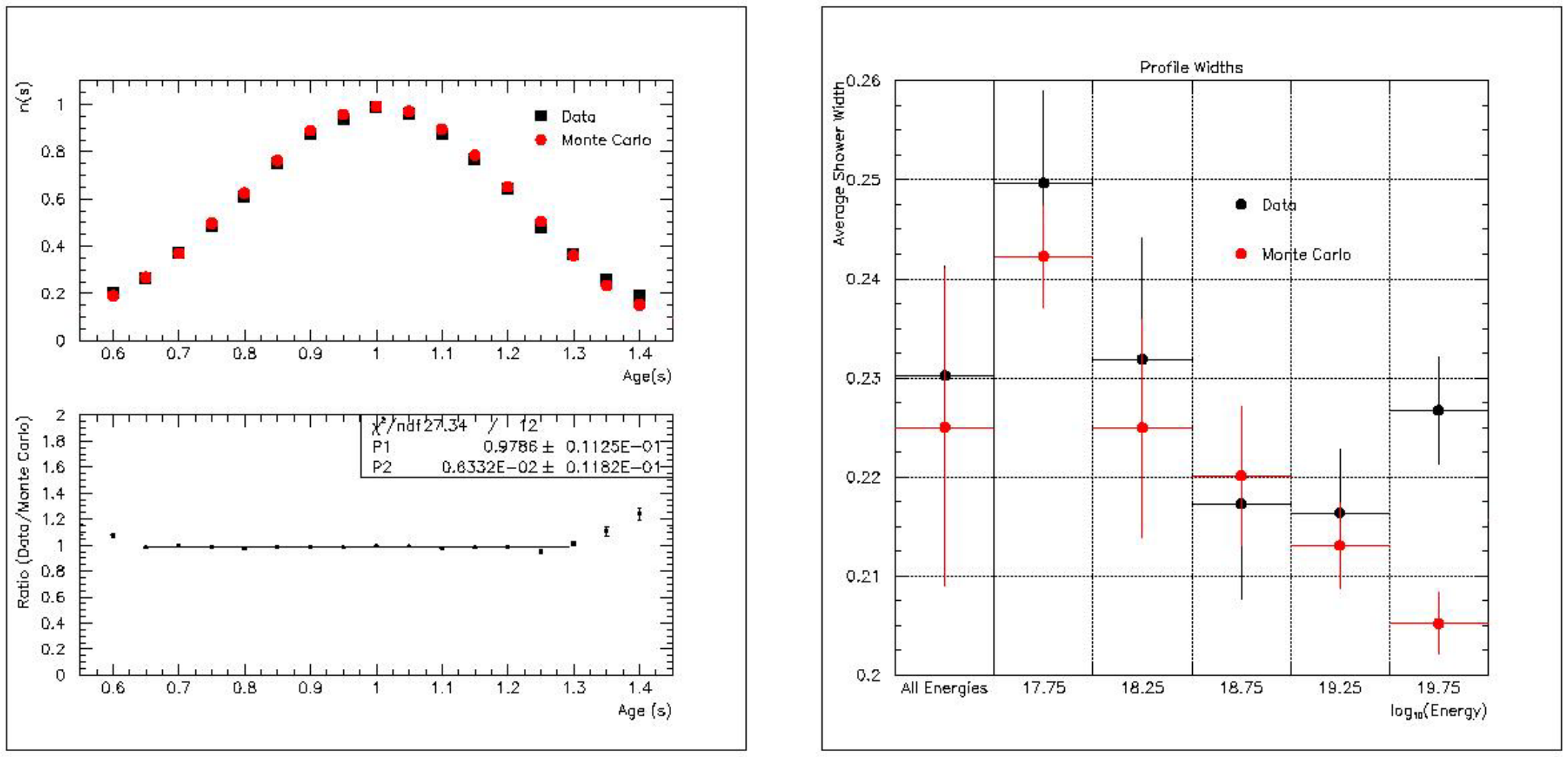

Figure 18. Preliminary plots of the longitudinal shower development profile for HiRes-II monocular data and Corsika/QGSjet simulations. The age distribution for E > $10^{18.5}$ eV is shown in the left hand plot, and a plot of data and Monte Carlo profile widths for half-decade bins in energy is shown in the right half plot. The agreement seems to be good, but there may be a discrepancy in shower profile widths for the highest energy bin.

**Conclusions**

One can look at the history of UHE cosmic ray experiments as having three phases. The original, pioneering phase, exemplified by experiments such as Akeno, Haverah Park, Yakutsk and Fly's Eye established the broad outlines of the subject, finding the first evidence of structure near the ankle and having just barely enough sensitivity to begin to explore the GZK region. In the second phase, the AGASA experiment gave us the first significant attempt to look for the GZK cutoff. Their observation of a continuing spectrum, together with the handful of events from earlier experiments above the GZK energy was a great puzzle. However, indirect methods of energy determination used by ground arrays can lead to misleading results. The HiRes experiment clearly established the existence of the GZK cutoff using the atmospheric fluorescence technique. The re-analysis of the AGASA data has decreased the discrepancy between it and the HiRes result and has re-iterated the importance of a better understanding of the meaning of ground array energy measurements. We are now entering the third phase, where ground array and air-fluorescence techniques are being combined in hybrid experiments such as the Telescope Array project in Utah and the Pierre Auger Observatory in Argentina. New information from the LHC and Fermilab experiments will allow further refinements in the modeling of EAS. Direct comparison of ground array and air-fluorescence data on an event-by-event basis and vis-à-vis simulations will improve knowledge of energy scales and experimental resolutions.

While it appears that the GZK puzzle has now been set to rest, the underlying questions are as urgent and burning as ever: what are the sources of UHE cosmic rays, what is the nature of the acceleration mechanism and what is the composition of the particles? Tantalizing hints of point sources have begun to appear in the last decade. Significant measurements of anisotropy, both on large and small scales, seem just within reach. The new generation of detectors, with order of magnitude larger sensitivities and excellent angular resolution should make a great impact in this area.

As the detailed structure of the cosmic ray spectrum becomes more reliably known, explicit fits to source models and propagation distances have become possible. Here experiments with a large dynamic range in energy are of the greatest importance, so that the region from the second knee to the GZK cutoff can be studied in a continuous fashion. The robust and reliable measurement of composition across this same energy range is also of the highest importance as it, together with broad-scale anisotropy, is the only way that the galactic and extra-galactic components of the UHE cosmic ray flux will be unraveled. Both the TA and Auger experiments are looking seriously at low-energy extensions to achieve these goals and the next decade should allow us to get a good understanding of the galactic-extragalactic transition region.

**References**

[1] S. Yoshida, et al., Astropart. Phys. 3, 105 (1995).
[2] Y. Arai, et al., Proc. 28th International Cosmic Ray Conference, Tsukuba, Japan Vol. 2 1025 (2003).
[3] J. Bluemer, Proc. 28th International Cosmic Ray Conference, Tsukuba, Japan HE 445 (2003).
[4] R.M. Baltrusaitis, et al., NIM, A240, 410 (1985).
[5] T. Abu-Zayyad, in Proceedings of the 25th International Cosmic Ray Conference, Durban, S. Africa, Vol. 5, 325 (1997); T. Abu-Zayyad et al., NIM A 450, 253 (2000).
[6] J. Linsley, Phys. Rev. Lett. 10, 146 (1963).
[7] M.M. Winn, et al., J. Phys. G 12, 653 (1986).
[8] M.A. Lawrence, R.J.O. Reid and A. A. Watson, J. Phys. G. 17, 773 (1991).
[9] N.N. Efimov, et al., in Astrophysical Aspects of the Most Energetic Cosmic Rays, ed. M. Nagano and F. Takahara (World Scientific) p. 20 (1991).
[10] M. Pagano, et al., J. Phys. G: Nucl. Phys. 18, 423 (1992).
[11] K. Greisen, Phys. Rev. Lett., 16 748 (1968); G. T. Zatsepin and V. A. K'uzmin, JETP Lett. 4, 78 (1966).
[12] V. Berezinsky, et al., hep-ph/0204357v3 (2007).
[13] V.S. Berezinsky and S. I. Grigor'evea, Astron. Astrophys. 199 1 (1988); V. S. Berezinsky, A.Z. Gazizov and S. I. Grigorieva, Phys Lett. B 612, 147 (2005).
[14] R.U. Abbasi, et al., Phys. Letters B 619, p. 271 (2005).
[15] T. Antoni, et al. Astropart. Phys. 24, 1-25 (2005).
[16] A.J. Bunner, et al., Canadian Journal of Physics 46, 5266 (1968).
[17] F. Kakimoto, et al., NIM Res A 372 527 (1996); M. Pagano, et al., Astropart. Phys. 22 235 (2004).
[18] J.W. Belz, et al, Astropart. Phys. in press (2006). P. Huentemyer, et al., AIP Conference Proceedings 698, 341 44 (2003).
[19] F. Arciprete, et al., Czechoslovak Journal of Physics, Vol. 56 Supl. 1, A361 (2006).
[20] R.U. Abbasi, et al., Astroparticle Physics, in press (2006). R.U. Abbasi, et al., submitted to Astroparticle Physics (2006).
[21] R. Clay et al. Proceedings of the 26th International Cosmic Ray Conference, Salt Lake City, USA, Vol. 5, p 421 (1999).
[22] T. Abu-Zayyad et al., NIM., A460, 278 (2001); T. Abu-Zayyad et al., Astropart. Phys. 18, 237 (2002).
[23] A.M. Hillas et al., Proc. 12 International Cosmic Ray Conference, Hobart, Tasmania, 3, 1001 (1971); A. M. Hillas, Acta Phys. Acad. Sci. Hung. 29, 355 (1970).
[24] A. J. Bower et al., J. Phys. G:Nucl Physics 9 L53 (1983).
[25] G.B. Christiansen et al., Proc. 19th International Cosmic Ray Conference, La Jolla, USA, 9 487 (1985).
[26] P. Mantsch et al., Proc. of the 29th International Cosmic Ray Conference, Puna, India, 00 101 (2005).
[27] J. Ranft, Phys. Rev. D 51 64 (1995).
[28] H.J. Drescher et al., Phys. Rep. 350 93 (2001).
[29] N.N. Kalmykov and S.S. Ostapchenko Phys. Atom Nucl. 56 346 (1993); N. N. Kalmykov et al. Nucl. Phys Proc. Suppl. 52B 17 (1997).
[30] R. Engel et al., Proc. 26th International Cosmic Ray Conference, Salt Lake City, Utah 1, 415 (1999).
[31] C. Meurer et al., Czech. J. Phys. 56 A211 (2006).
[32] H. J. Drescher et al., Astropart. Phys. 21 87 (2004).
[33] H. Meyer, Nuclear Physics B – Proceedings Supplements, 142 453; S. Seun, Nuclear Physics B – Proc. Supp. 142 566

[34] M. Deile arxiv/hep-ex/0410084; B. E. Cox, hep-ph/0609209; O. Adriani et al., Czechoslovak Journal of Physics, Vol 56, Supl. 1, 2006, **A107.**

[35] K. Werner et al., Phys. Rev. C 74, 044902 (2006).

[36] T.Pierog and K. Werner, astro-ph/0611311 (2006).

[37] D.J. Bird *et al.*, Phys. Rev. Lett. **71**, 3401 (1993).

[38] D.J. Bird *et al.*, Ap. J. **441**, 144 (1995).

[39] J. Linsley, Phys. Rev. Lett. **10**, 146 (1963); M.A. Lawrence *et al.*, J. Phys. G. **17** 733 (1991).

[40] N. Hayashida et al, Ap. J. 522, 225 (1999); M. Takeda *et al*., Astropart. Phys. **19** 447 (2003).

[41] K. Shinozaki, Proceedings, Quarks2006 Conference, Repino, Russia; M. Teshima et al., Proceedings of CRIS2006, Catania, Sicily (2006).

[42] D. Heck, Forschungszentrum Karlsruhe Report FZKA 6019, (1998).

[43] S.J. Sciutto astro-ph/9911331.

[44] R.U. Abbasi et al.), Phys. Lett. B 619, p. 271 (2005).

[45] P. Sokolsky et al., Proceedings of CRIS2006, Catania, Sicily (2006).

[46] T.K. Gaisser et al. Comments on Astrophysics, Vol.17, No.2 & 3, 103 (1993); T.K. Gaisser et al. Phys. Rev. D, 47, 1919 (1993).

[47] D. Allard et al., Astropart. Phys. 27, 61 (2007).

[48] R.U. Abbasi et al., Ap. J. 622 910 (2005).

[49] G.B. Thomson et al.) Proceedings of the 29th International Cosmic Ray Conference, Pune, India Vol. (2005); G. B. Thomson et al. Proceedings of the 29th International Cosmic Ray Conference, Pune, India, Vol. (2005).

[50] R.U. Abbasi et al. in Proceedings of the 29th International Cosmic Ray Conference (ICRC 2005), Pune, India.

[51] N. Hayashida et al., J. Phys. G: Nucl. Part. Phys. 21 1101 (1995).

[52] T. Abu-Zayyad et al., Phys. Rev. Lett, 84 (2000) p. 4276.

[53] H. Hayashida et al., (1995) op cit.

[54] M. Nagano et al., Astropart. Phys. 13, 4 277 (2000).

[55] K. Werner et al., Phys. Rev. C 74, 044902 (2006).

[56] J. A. Goodman et al., Phys. Rev. D 26, 1043 (1982).

[57] M. Honda *et al*, Phys. Rev. Lett. **70** 525. (1993).

[58] R.M. Baltrusaitis et al., Phys. Rev. Lett. 52, 1380 (1984).

[59] R.U. Abbasi et al. in Proceedings of the 29th International Cosmic Ray Conference (ICRC 2005), Pune, India.

[60] R. W. Ellsworth et al., Phys Rev. D 26 336 (1982).

[61] R.D. Blandford and J. P. Ostriker, Astrophysical J. 237, 793 (1980); D. C. Ellison et al., Astrophys. J. 540, 292 (2000).

[62] J.A. Hinton New Astron. Rev. 48 331 (2004); F. Aharonian et al., Astron. Astrophys. 430, 865 (2005).

[63] J.C. Higdon et al, Astrophysical J. 509, 1, L33 (1998); W. R. Binns et al, Proceedings of the Workshop on the End of the Galactic Cosmic Ray Spectrum, Aspen, (2005).

[64] F. Melia and H. Falcke, Annu. Rev. Astron. Astrophys. 39, 309 (2001).

[65] M. Takeda et al., Astropart. Phys. 19, 447 (2003).

[66] R.U. Abbassi et al., Ap. J., 610 L79 (2004); R. U. Abbasi et al., Astropart. Phys., 22 139 (2004).

[67] R. U. Abbasi et al., Ap. J., 623 164 (2005).

[68] I. Tkatchev et al., JETP **74**, 445 (2001)

[69] D. Gorbunov et al., Ap. J. 577, L93 (2002)]

[70] D. Gorbunov et al., JETP Lett. **80**, 145 (2004).

[71] R.U. Abbasi et al., Ap. J., 636 680 (2006).

[72] N. Hayashida et al., Astropart. Phys. 10 303 (1999).

[73] M. Aglieta et al., astro-ph/0607382 (2006).

[74] T. Stanev. Ap. J. 479 part 1, 290 (1979).
[75] T. Abu-Zayyad et al., Astropart. Phys., 16 1 (2001).